\documentclass[preprint,sort&compress,10pt]{elsarticle}

\usepackage{graphicx}

\usepackage{amssymb}
\usepackage{bbm}
\usepackage{amsmath}
\usepackage{color}
\usepackage[letterpaper]{geometry}
\journal{International Journal of Solids and Structures}

\begin{document}

\begin{frontmatter}



\title{\textbf{Structure of Defective Crystals at Finite Temperatures: \\A Quasi-Harmonic Lattice Dynamics Approach}}


\author{Arash Yavari and Arzhang Angoshtari}

\address{School of Civil and Environmental Engineering, Georgia Institute of Technology, \\ Atlanta, GA 30332, U.S.A.}

\begin{abstract}
In this paper we extend the classical method of lattice dynamics
to defective crystals with partial symmetries. We start by a
nominal defect configuration and first relax it statically. Having
the static equilibrium configuration, we use a quasiharmonic lattice
dynamics approach to approximate the free energy. Finally, the
defect structure at a finite temperature is obtained by minimizing
the approximate Helmholtz free energy. For higher temperatures we
take the relaxed configuration at a lower temperature as the
reference configuration. This method can be used to semi-analytically study the
structure of defects at low but non-zero temperatures, where molecular dynamics cannot be used. As an
example, we obtain the finite temperature structure of
two $180^{\circ}$ domain walls in a 2-D lattice of interacting dipoles.
We dynamically relax both the position and polarization vectors.
In particular, we show that increasing temperature the
domain wall thicknesses increase.
\end{abstract}


\end{frontmatter}

\section{\textbf{Introduction}}

Although it has been recognized that defects play an important
role in nanostructured materials, the fundamental understanding of
how defects alter the material properties is not satisfactory. The
link between defects and the macroscopic behavior of materials
remains a challenging problem. Classical mechanics of defects that
studies materials with microscale defects is based on continuum
theories with phenomenological constitutive relations. In the
nanoscale, the continuum quantities such as stress and strain
become ill defined. In addition, due to size effects, to study
defects in nano-structured materials, non-classical solutions of
defect fields is necessary~\citep{Gutkin2006}. The application of
continuum mechanics to small-scale problems is problematic;
atomistic numerical methods such as ab initio
calculations~\citep{MeyerVanderbilt2001,Ogata2009},
Molecular Dynamics (MD) simulations~\citep{Jang2007,Guo2005} and
Monte Carlo (MC) simulations~\citep{Zetterstrom2005,Mok2007} can
be used for nanoscale mechanical analyses. However, the
application of these methods is largely restricted by the size
limit and the periodicity requirements. Current ab initio
techniques are unable of handling systems with more than a few
hundred atoms. Molecular dynamics simulations can model larger
systems, however, MD is based on equations of classical mechanics
and thus cannot be used for low temperatures, where quantum
effects are dominant. Engineering with very small structures
requires the ability to solve inverse problems and this cannot be
achieved through purely numerical methods. What is ideally needed
is a systematic method of analysis of solids with defects that is
capable of treating finite temperature effects.

The only analytic/semi-analytic method for solving
zero-temperature defect problems in the lattice scale is the
method of lattice statics. The method of lattice statics was
introduced in \citep{Matsubara1952,Kanazaki1957}. This method has been
used for point defects \citep{FlockenHardy1969,Flocken72}, for
cracks \citep{Esterling1978a,Esterling1978b,HsiehThomson1973},
and also for dislocations \citep{BoyerHardy1971,Esterling1978b,EsterlingMoriarty1978,Maradudin1958,Shenoy1999,Tewary2000}.
More details and history can be found in
\citep{BornHuang1998,BoyerHardy1971,Bull79,FlockenHardy1969,FlockenHardy1970,Gall93,Mar71,Ortiz1999,Shenoy1999,Tew73}
and references therein. Lattice statics is based on
energy minimization and cannot be used at finite temperatures. The
other restriction of most lattice statics calculations is the
harmonic approximation, which can be too crude close to defects.
Recently, motivated by applications in ferroelectrics, we developed a general theory of
anharmonic lattice statics capable of semi-analytic modeling of
different defective crystals governed by different types of
interatomic potentials \citep{Yavari2005a,Yavari2005b,KavianpourYavari2009}. At finite
temperatures, the use of quantum mechanics-based lattice dynamics
is necessary. Unfortunately, lattice dynamics has mostly been used
for perfect crystals and for understanding their thermodynamic
properties
\citep{BornHuang1998,Dove1993,Kittel1987,Kossevich1999,Mar71,Peierls1955,Wallace1965}.
There is not much in the literature on corrections for anharmonic
effects and systematic solution techniques for defective crystals.
Some of these issues will be addressed in this paper.

In order to accurately predict the mechanical properties of nanosize
devices one would need to take into account the effect of finite
temperatures. It should be mentioned that most multiscale methods so
far have been formulated for $T=0$ calculations. An example is the
quasi-continuum method \citep{Ortiz1999,Tadmor1996}. However,
recently there have been several attempts in extending this method
for finite temperatures
\citep{DiestlerWuZeng2004,Dupuy2005,KulkarniKnapOrtiz2008,Tang2006}.
As Forsblom, et al. \citep{ForsblomSandbergGrimvall2004} mention,
very little is known about the vibrational properties of defects in
crystalline solids. Sanati and Esetreicher
\citep{SanatiEsetreicher2003} showed the importance of vibrational
effects in semi-conductors and the necessity of free energy
calculations. Lattice dynamics \citep{BornHuang1998,Peierls1955} has
been ignored with the exception of some very recent works
\citep{TaylorBarreraAllanBarron1997}. As examples of
finite-temperature defect solutions we can mention Taylor, et al.
\citep{TaylorAllanBrunoBarrera1999,TaylorBarreraAllanBarronMackrodt1997}
who discuss quasiharmonic lattice dynamics for three-body
interactions in bulk crystals. Taylor, et al.
\citep{TaylorSimsBarreraAllanMackrodt1999} consider a slab, i.e., a
system that is periodic only in two directions. They basically
consider a supercell that is repeated in the plane periodically. As
Allan, et al.
\citep{AllanBarreraPurtonSimsTaylor2000,AllanBarronBruno1996}
conclude, a combination of quasiharmonic lattice dynamics, molecular
dynamics, Monte Carlo simulations and ab initio calculations should
be used in real applications. However, at this time there is no
systematic method of lattice dynamics for thermodynamic analysis of
defective systems that is also capable of capturing the anharmonic
effects. We should mention that in many materials systems lattice
dynamics is a valid approximation up to two-third of the bulk
melting temperature but it turns out that harmonic approximation may
not be adequate for free energy calculations of defects at high
temperatures (see \citep {Foiles1994} for discussions on Cu).
Hansen, et al. \citep{HansenVoglFiorentini1999} show that for Al
surfaces above the Debye temperature quasiharmonic lattice dynamic
approximation starts to fail. Zhao, et al.
\citep{ZhaoTangLiAluru2006} show that quasiharmonic lattice dynamics
accurately predicts the thermodynamic properties of silicon for
temperatures up to $800~K$. In this paper, we are interested in low
temperatures where MD fails while quasi-harmonic lattice dynamics is
a good approximation.

For understanding defect structures the main quantity of interest is the Helmholtz free energy. Free energy is an important thermodynamic function that determines
the relative phase stability and can be used to generate other
thermodynamic functions. In \emph{quasiharmonic}
lattice dynamics, for a system of $n$ atoms, free energy is
computed by diagonalizing a $3n\times 3n$ matrix that is obtained
by quadratizing the Hamiltonian about a given static equilibrium
configuration. Using similar ideas, for a perfect crystal with a
unit cell with $N$ atoms, one can compute the free energy by
diagonalizing a $3N\times 3N$ matrix in the reciprocal space. In
the \emph{local quasiharmonic} approximation one assumes that
atoms vibrate independently and thus all is needed for calculation
of free energy is to diagonalize $n$ $3\times 3$ matrices
\citep{LesarNajafabadiSrolovitz1989} (see Rickman and LeSar
\citep{RickmanLeSar2002} for a recent review of the existing
methods for free energy calculations). These will be discussed in more detail in \S3.

In this paper, we propose a theoretical framework of quasi-harmonic
lattice dynamics to address the mechanics of defects in crystalline
solids at low but finite temperatures. The main ideas are summarized
as follows. We think of a defective lattice problem as a discrete
deformation of a collection of atoms to a discrete current
configuration. The lattice atoms are assumed to interact through
some interatomic potentials. At finite temperatures, the equilibrium
positions of the atoms are not the same as their static equilibrium
($T=0$) positions; the lattice atoms undergo thermal vibrations.
The potential and Helmholtz free energies of the lattice are taken
as discrete functionals of the discrete deformation mapping. For
finite temperature equilibrium problems, the discrete nonlinear
governing equations are linearized about a reference configuration.
The finite-temperature equilibrium configuration of the defective
lattice can then be obtained semi-analytically. For finite
temperature dynamic problems, the Euler-Lagrange equations of motion
of the lattice are casted into a system of ordinary differential
equations by superimposing the phonon modes. We should emphasize
that our method of lattice dynamics is not restricted to finite
systems; defects in infinite lattices can be analyzed
semi-analytically. The only restriction is the use of interatomic
potentials.

This paper is structured as follows. In \S 2 we briefly review the
theory of anharmonic lattice statics presented in
\citep{Yavari2005a} and \citep{Yavari2005b}. We then present an
overview of the basic ideas of the method of lattice dynamics for
both finite and infinite atomic systems in \S 3. This follows by an extension of
these ideas to defective crystals with partial symmetries. In \S 4
we formulate the lattice dynamics governing equations for a
2-D lattice of dipoles with both short and long-range interactions. In
\S 5 we study the temperature dependence of the structure of
two $180^{\circ}$ domain walls in the dipole lattice.
Conclusions are given in \S 6.

\section{\textbf{Anharmonic Lattice Statics}}

Consider a collection of atoms $\mathcal{L}$ with the current
configuration
$\left\{\mathbf{x}^i\right\}_{i\in\mathcal{L}}\subset\mathbb{R}^n$.
Assuming that there is a discrete field of body forces
$\{\mathbf{F}^i\}_{i\in\mathcal{L}}$, a necessary condition for
the current position $\{\mathbf{x}^i\}_{i\in\mathcal{L}}$ to be in
static equilibrium is $-\frac{\partial \mathcal{E}}{\partial
   \mathbf{x}^i}+\mathbf{F}^i=\mathbf{0},~\forall~i\in\mathcal{L}$,  where $\mathcal{E}$ is the total static energy and is a function
of the atomic positions. These discrete governing equations are
highly nonlinear. In order to obtain semi-analytical solutions, we
first linearize the governing equations with respect to a
reference configuration
$\mathcal{B}_0=\{\mathbf{x}^i_0\}_{i\in\mathcal{L}}$
\citep{Yavari2005a}. We leave the reference configuration
unspecified; at this point it would be enough to know that we
usually choose the reference configuration to be a nominal defect
configuration \citep{Yavari2005a, Yavari2005b,
KavianpourYavari2009}.

Taylor expansion of the governing equations for an atom $i$ about
the reference configuration
$\mathcal{B}_0=\{\mathbf{x}^i_0\}_{i\in\mathcal{L}}$ reads
\begin{equation}
    -\frac{\partial \mathcal{E}}{\partial \mathbf{x}^i}+\mathbf{F}^i= -\frac{\partial \mathcal{E}}{\partial \mathbf{x}^i}\left(\mathcal{B}_0\right)-
    \frac{\partial^2 \mathcal{E}}{\partial \mathbf{x}^i\partial \mathbf{x}^i} \left(\mathcal{B}_0\right)\cdot(\mathbf{x}^i-\mathbf{x}^i_0)
  -\sum_{\begin{subarray}{l} j\in\mathcal{L}\\ j\neq i \end{subarray}}\frac{\partial^2 \mathcal{E}}{\partial \mathbf{x}^j\partial
  \mathbf{x}^i}\left(\mathcal{B}_0\right)\cdot(\mathbf{x}^j-\mathbf{x}^j_0)- ...+\mathbf{F}^i=\mathbf{0}.
\end{equation}
Ignoring terms that are quadratic and higher in
$\{\mathbf{x}^j-\mathbf{x}^j_0\}$, we obtain
\begin{equation}
    \frac{\partial^2 \mathcal{E}}{\partial \mathbf{x}^i\partial
    \mathbf{x}^i}\left(\mathcal{B}_0\right)\cdot(\mathbf{x}^i-\mathbf{x}^i_0)
    +\sum_{\begin{subarray}{l} j\in\mathcal{L}\\ j\neq i \end{subarray}}\frac{\partial^2 \mathcal{E}}{\partial \mathbf{x}^j\partial
    \mathbf{x}^i}\left(\mathcal{B}_0\right)\cdot(\mathbf{x}^j-\mathbf{x}^j_0)=-\frac{\partial
    \mathcal{E}}{\partial \mathbf{x}^i}\left(\mathcal{B}_0\right)+\mathbf{F}^i~~~~~\forall i\in \mathcal{L}.
\end{equation}
Here, $\left\{-\frac{\partial \mathcal{E}}{\partial
\mathbf{x}^i}\left(\mathcal{B}_0\right)\right\}_{i\in\mathcal{L}}$
is the discrete field of unbalanced forces.

\paragraph{Defective Crystals and Symmetry Reduction.} In many defective crystals one can simplify the calculations
by exploiting symmetries. A defect, by definition, is anything
that breaks the translation invariance symmetry of the crystal.
However, it may happen that a given defect does not affect the
translation invariance of the crystal in one or two directions.
With this idea, one can classify defective crystals into three
groups: (i) with 1-D symmetry reduction, (ii) with 2-D symmetry
reduction and (iii) with no symmetry reduction. Examples of (i),
(ii) and (iii) are free surfaces, dislocations, and point defects,
respectively \citep{Yavari2005a}. Assume that the defective
crystal $\mathcal{L}$ has a 1-D symmetry reduction, i.e. it can be
partitioned into two-dimensional equivalence classes as follows
\begin{equation}
    \mathcal{L}=\bigsqcup_{\alpha\in\mathbb{Z}}\bigsqcup_{I=1}^{N}\mathcal{S}_{I\alpha},
\end{equation}
where $\mathcal{S}_{I\alpha}$ is the equivalence class of all the
atoms of type $I$ and index $\alpha$ (see \citep{Yavari2005a} and
\citep{KavianpourYavari2009} for more details). Here, we assume
that $\mathcal{L}$ is a multilattice of $N$ simple lattices. For a
free surface, for example, each equivalence class is a set of
atoms lying on a plane parallel to the free surface. Using this
partitioning for $i=I\alpha$ one can write
\begin{equation}
    \sum_{\begin{subarray}{l} j\in\mathcal{L}\\ j\neq i \end{subarray}}\frac{\partial^2 \mathcal{E}}{\partial \mathbf{x}^j\partial
    \mathbf{x}^i}\left(\mathcal{B}_0\right)\cdot(\mathbf{x}^j-\mathbf{x}^j_0)
    =\sideset{}{'}{\sum}_{\beta\in \mathbb{Z}}\sum_{J=1}^N\sum_{j \in \mathcal{S}_{J\beta}}
    \frac{\partial^2 \mathcal{E}}{\partial \mathbf{x}^j \partial \mathbf{x}^{i}}(\mathcal{B}_0)
    \cdot\left(\mathbf{x}^{J\beta}-\mathbf{x}^{J\beta}_0\right),
\end{equation}
where the prime on the first sum means that the term
$J\beta=I\alpha$ is omitted. The linearized discrete governing
equations are then written as \citep{Yavari2005a}
\begin{equation} \label{equilibrium}
    \sideset{}{'}{\sum}_{\beta\in \mathbb{Z}}\sum_{J=1}^{N} \mathbf{K}_{I\alpha J\beta}
    \mathbf{u}^{J\beta}+
    \left(-\sideset{}{'}{\sum}_{\beta\in \mathbb{Z}}\sum_{J=1}^{N} \mathbf{K}_{I\alpha J\beta}\right)\mathbf{u}^{I\alpha}=
    \mathbf{f}_{I\alpha},
\end{equation}
where
\begin{equation}
  \mathbf{K}_{I\alpha J\beta}=\sum_{j \in \mathcal{S}_{J\beta}}\frac{\partial^2 \mathcal{E}}{\partial
  \mathbf{x}^j \partial \mathbf{x}^{I\alpha}}(\mathcal{B}_0), ~~ \mathbf{f}_{I\alpha}
  =-\frac{\partial \mathcal{E}}{\partial \mathbf{x}^{I\alpha}}(\mathcal{B}_0)+\mathbf{F}_{I\alpha},~~  \mathbf{u}^{J\beta}=\mathbf{x}^{J\beta}-\mathbf{x}^{J\beta}_0=\mathbf{x}^{j}-\mathbf{x}^{j}_0
  ~~~~~ \forall ~j \in \mathcal{S}_{J\beta}.
\end{equation}
The governing equations in terms of unit cell displacement vector
$\mathbf{U}_{\alpha}=\left(\mathbf{u}^{1}_{\alpha},...,\mathbf{u}^{N}_{\alpha}\right)^{\textsf{T}}$
can be written as
\begin{equation}\label{difference-1-D}
    \sum_{\beta\in \mathbb{Z}}\mathbf{A}_{\beta}(\alpha)\mathbf{U}_{\alpha+\beta}
    =\mathbf{F}_{\alpha}~~~~~\alpha\in\mathbb{Z},
\end{equation}
where $\mathbf{A}_{\beta}(\alpha)\in\mathbb{R}^{3N\times3N},~\mathbf{U}_{\alpha},\mathbf{F}_{\alpha}\in\mathbb{R}^{3N}$. This is a linear vector-valued ordinary difference equation with
variable coefficient matrices. The unit cell force vectors and the
unit cell stiffness matrices are defined as
\begin{equation}
    \mathbf{F}_{\alpha}=\left(
                       \begin{array}{c}
                         \mathbf{F}_{1\alpha} \\
                         \vdots \\
                         \mathbf{F}_{N\alpha} \\
                       \end{array}
                     \right),~
    \mathbf{A}_{\beta}(\alpha)=\left(
                                            \begin{array}{cccc}
                                              \mathbf{K}_{1\alpha 1\beta} & \mathbf{K}_{1\alpha 2\beta} & \cdots & \mathbf{K}_{1\alpha N\beta} \\
                                              \mathbf{K}_{2\alpha 1\beta} & \mathbf{K}_{2\alpha 2\beta} & \cdots & \mathbf{K}_{2\alpha N\beta} \\
                                              \vdots & \vdots & \cdots & \vdots \\
                                              \mathbf{K}_{N\alpha 1\beta} & \mathbf{K}_{N\alpha 2\beta} & \cdots & \mathbf{K}_{N\alpha N\beta} \\
                                            \end{array}
                                          \right)~~~~~\alpha,\beta\in\mathbb{Z}.
\end{equation}
Note that, in general, $\mathbf{A}_{\beta}$ need not be symmetric
\citep{Yavari2005a}. The resulting system of difference equations can be solved directly or using discrete Fourier transform \citep{Yavari2005a}.

\paragraph{Hessian Matrix for the Bulk Crystal.} A
bulk crystal is a defective crystal with a $0$-D symmetry
reduction. Governing equations for atom $I$ in the unit cell
$\mathbf{n}=\mathbf{0}$ read $-\frac{\partial
    \mathcal{E}}{\partial\mathbf{x}^{I}}+\mathbf{F}_I=\mathbf{0},~I=1,...,N$. Linearization about $\mathcal{B}_0=\{\mathbf{X}^I\}$ yields
\begin{equation}
    \frac{\partial^2\mathcal{E}}{\partial\mathbf{x}^{I}\partial\mathbf{x}^{I}}(\mathcal{B}_0)\cdot(\mathbf{x}^I-\mathbf{X}^I)
    +\sum_{\begin{subarray}{l} j\in\mathcal{L}\\ j\neq I \end{subarray}}\frac{\partial^2\mathcal{E}}{\partial\mathbf{x}^I\partial\mathbf{x}^j}
    (\mathcal{B}_0)\cdot(\mathbf{x}^j-\mathbf{X}^j)+...=-\frac{\partial
    \mathcal{E}}{\partial\mathbf{x}^I}(\mathcal{B}_0)+\mathbf{F}_I~~~~~~~I=1,...,N.
\end{equation}
Note that
\begin{equation}
    \sum_{\begin{subarray}{l} j\in\mathcal{L}\\ j\neq I \end{subarray}}\frac{\partial^2\mathcal{E}}{\partial\mathbf{x}^I\partial\mathbf{x}^j}
    (\mathcal{B}_0)\cdot(\mathbf{x}^j-\mathbf{X}^j)=\sum_{\begin{subarray}{l} J=1\\ J\neq I \end{subarray}}^{N}\sum_{j\in\mathcal{L}_J}
    \frac{\partial^2\mathcal{E}}{\partial\mathbf{x}^I\partial\mathbf{x}^j}
    (\mathcal{B}_0)\cdot(\mathbf{x}^j-\mathbf{X}^j)
    +\sum_{\begin{subarray}{l} j\in\mathcal{L}_I\\ j\neq I \end{subarray}}
    \frac{\partial^2\mathcal{E}}{\partial\mathbf{x}^I\partial\mathbf{x}^j}
    (\mathcal{B}_0)\cdot(\mathbf{x}^j-\mathbf{X}^j).
\end{equation}
We also know that because of translation invariance of the
potential
\begin{equation}
    \frac{\partial^2\mathcal{E}}{\partial\mathbf{x}^{I}\partial\mathbf{x}^{I}}(\mathcal{B}_0)=
    -\sum_{\begin{subarray}{l} j\in\mathcal{L}\\ j\neq I \end{subarray}}
    \frac{\partial^2\mathcal{E}}{\partial\mathbf{x}^I\partial\mathbf{x}^j}
    (\mathcal{B}_0).
\end{equation}
Therefore, the linearized governing equations can be written as
\begin{equation}
    \sum_{\begin{subarray}{l} J=1\\ J\neq I \end{subarray}}^{N}\mathbf{K}_{IJ}\mathbf{u}^{J}
    +\Bigg(\!\!-\sum_{\begin{subarray}{l} J=1\\ J\neq I
    \end{subarray}}^{N}\mathbf{K}_{IJ}\Bigg)
    \mathbf{u}^I=\mathbf{f}^I~~~~~~~~~I=1,...,N,
\end{equation}
where
\begin{equation}
    \mathbf{K}_{IJ}=\sum_{j\in\mathcal{L}_J}
    \frac{\partial^2\mathcal{E}}{\partial\mathbf{x}^I\partial\mathbf{x}^j}
    (\mathcal{B}_0),~~~
   \mathbf{f}^I=-\frac{\partial \mathcal{E}}{\partial\mathbf{x}^I}(\mathcal{B}_0)+\mathbf{F}_I,~~~
   \mathbf{u}^J=\mathbf{x}^J-\mathbf{X}^J=\mathbf{x}^j-\mathbf{X}^j~~~~~\forall~j\in\mathcal{L}_J.
\end{equation}
The Hessian matrix of the bulk crystal is defined as
\begin{equation}
    \mathbf{H}=
    \left(%
    \begin{array}{cccc}
      \mathbf{K}_{11} & \mathbf{K}_{12} & \ldots & \mathbf{K}_{1N} \\
      \mathbf{K}_{21} & \mathbf{K}_{22} & \ldots & \mathbf{K}_{2N} \\
      \vdots & \vdots & \ddots & \vdots \\
      \mathbf{K}_{N1} & \mathbf{K}_{N2} & \ldots & \mathbf{K}_{NN} \\
    \end{array}%
    \right),
\end{equation}
where $\mathbf{K}_{JI}=\mathbf{K}_{IJ}$. Stability of the bulk
crystal dictates $\mathbf{H}$ to be positive-semidefinite with
three zero eigenvalues. In the case of a defective crystal, one can look at a sequence of sublattices containing the defect and calculate the corresponding sequence of Hessians.

\section{Method of Quasi-Harmonic Lattice Dynamics}

At a finite temperature $T$ (constant volume) thermodynamic
stability is governed by Helmholtz free energy $F=E-TS$. In
principle, $F$ is well-defined in the setting of statistical
mechanics. Quantum-mechanically calculated energy levels $E(i)$ for
different microscopic states can be used to obtain the partition
function \citep{KittelKroemer1980,Weiner2002}
\begin{equation}
    Q=\sum_{i}\exp\left(\frac{-E(i)}{k_BT}\right),
\end{equation}
where $k_B$ is Boltzman's constant. Finally $F=-k_BT\ln Q$ (see the
appendix). However, one should note that the phase space is
astronomically large even for a finite system. Usually, in practical
problems, molecular dynamics and Monte Carlo simulations, coupled
with thermodynamic integration techniques, reduce the complexity of
the free energy calculations. For low to moderately high
temperatures, quantum treatment of lattice vibrations in the
harmonic approximation provides a reliable description of
thermodynamic properties \citep{Mar71}. In the following we review
the classical formulation of lattice dynamics first for a finite
collection of atoms and then for bulk crystals.

\subsection{Finite Systems}

For a finite system of $N$ atoms suppose $\mathcal{B} =
\left\{\mathbf{X}^i\right\}_{i\in \mathcal{L}}$ is the static
equilibrium configuration, i.e. $\frac{\partial \mathcal{E}}{\partial \mathbf{x}^i}\big|_{\mathbf{x}^i=\mathbf{X}^i}=\mathbf{0},~\forall~i \in \mathcal{L}$. Hamiltonian of this collection is written as
\begin{equation}
    \mathcal{H}\left(\left\{\mathbf{x}^i\right\}_{i\in \mathcal{L}}\right)=\frac{1}{2}\sum_{i\in\mathcal{L}}m_i|\dot{\mathbf{x}}^i|^2
    +\mathcal{E}\left(\left\{\mathbf{x}^i\right\}_{i\in \mathcal{L}}\right).
\end{equation}
Now denoting the thermal displacements by
$\mathbf{u}^i=\mathbf{x}^i-\mathbf{X}^i$ potential energy of the system is written as
\begin{equation}
    \mathcal{E}\left(\left\{\mathbf{x}^i\right\}_{i\in \mathcal{L}}\right)=\mathcal{E}\left(\left\{\mathbf{X}^i\right\}_{i\in \mathcal{L}}\right)
    +\frac{1}{2}\sum_{i,j\in\mathcal{L}}\mathbf{u}^i{}^{\textsf{T}}\cdot\frac{\partial^2\mathcal{E}}{\partial\mathbf{x}^i\partial\mathbf{x}^j}(\mathcal{B})\mathbf{u}^j+....
\end{equation}
Or
\begin{equation}
    \mathcal{E}(\mathbf{x})=\mathcal{E}(\mathbf{X})+\frac{1}{2}\mathbf{u}^{\textsf{T}}\boldsymbol{\Phi}\mathbf{u}+o(|\mathbf{u}|^2),
\end{equation}
where $\boldsymbol{\Phi}$ is the matrix of force constants. The
Hamiltonian is approximated by
\begin{equation}
    \mathcal{H}(\mathbf{x})=\mathcal{E}(\mathbf{X})+\frac{1}{2}\mathbf{u}^{\textsf{T}}\boldsymbol{\Phi}\mathbf{u}
    +\frac{1}{2}\dot{\mathbf{u}}^{\textsf{T}}\mathbf{M}\dot{\mathbf{u}},
\end{equation}
where $\mathbf{M}$ is the diagonal mass matrix. Let us denote the
matrix of eigenvectors of $\boldsymbol{\Phi}$ by $\boldsymbol{U}$,
and write
\begin{equation}
    \mathcal{H}(\mathbf{x})=\mathcal{E}(\mathbf{X})+\frac{1}{2}\mathbf{q}^{\textsf{T}}\boldsymbol{\Lambda}\mathbf{q}
    +\frac{1}{2}\dot{\mathbf{q}}^{\textsf{T}}\mathbf{M}\dot{\mathbf{q}},
\end{equation}
where $\mathbf{q}=\mathbf{U}^{\textsf{T}}\mathbf{u}$ is the vector
of normal displacements and
$\boldsymbol{\Lambda}=\textrm{diag}(\lambda_1,...,\lambda_{3N})$
is the diagonal matrix of eigenvalues of $\boldsymbol{\Phi}$. This
is now a set of $3N$ independent harmonic oscillators. Solving
Schr\"{o}ndinger's equation gives the energy levels of the
r\emph{th} oscillator as \citep{Mar71}
\begin{equation}\label{Frequencies}
    E_{nr}=\mathcal{E}_r(\mathbf{X})+\left(n+\frac{1}{2}\right)\hbar\omega_r~~~~~n=0,1,...,~r=1,...,3N,
\end{equation}
where $\omega_r=\omega_r\left(\left\{\mathbf{X}^i\right\}_{i\in
\mathcal{L}}\right)=\sqrt{\lambda_r/m_r}$. The free energy is then
written as \citep{BornHuang1998}
\begin{eqnarray}
  \mathcal{F}\left(\left\{\mathbf{X}^i\right\}_{i\in \mathcal{L}}, T\right) &=& -k_BT\sum_{r=1}^{3N}\ln\sum_{n=0}^{\infty}\exp\left(\frac{-E_{nr}}{k_BT}\right) \nonumber \\
   &=&
   \mathcal{E}\left(\left\{\mathbf{X}^i\right\}_{i\in \mathcal{L}}\right)
   +\frac{1}{2}\sum_{r=1}^{3N}\hbar\omega_r+k_BT\sum_{r=1}^{3N}\ln\left[1-\exp \left(-\frac{\hbar\omega_r}{k_BT}\right)\right].
\end{eqnarray}

Here it should be noted that we have considered a time-independent
Hamiltonian, which can be regarded as a first-order approximation
for some problems. Assume that Hamiltonian $H$ of a system contains a time-dependent parameter $\mathbf{f}(t)$, say a
time-dependent external force. If the time variation of
$\mathbf{f}\left(t\right)$ is slow and does not cause a large
variation of $H$ in a time interval of the same order as the
natural period of the system with constant $\mathbf{f}$, then this
approximation is valid \citep{Nogami1991}, otherwise one should
consider time-dependent harmonic oscillator systems. This can be
the case for various quantum mechanical systems
\citep{KiwiRossler1972,LimaRosasPedrosa2008,Meyer1981}. In such
situations one should obtain the solution of Schr\"{o}ndinger's
equation for a time-dependent forced harmonic oscillator and as
a result, energy levels would depend on the forcing terms too. As an
example, Meyer \citep{Meyer1981} investigated energy propagation
in a one-dimensional finite lattice with a time-dependent
driving forces by solving the corresponding forced Schr\"{o}ndinger's equation. We also mention that the above
formula for the free energy is based on the quasiharmonic
approximation. As temperature increases such an approximation may
become invalid for some materials \citep{LacksRutledge1994} and
therefore one would need to consider anharmonic effects. To
include anharmonic terms in the free energy relation, anharmonic
perturbation theory can be used by choosing the quasiharmonic
state as the unperturbed state and the perturbation is due to the
terms higher than second order in the Taylor expansion of the
potential energy \citep{ShuklaCowley1971}. This way, one accounts
for anharmonic coupling of the vibrational modes.

As we discuss in the appendix, to obtain the optimum positions of
atoms at a constant temperature $T$ one should minimize the free
energy with respect to all the geometrical variables
$\left\{\mathbf{X}^i\right\}_{i\in \mathcal{L}}$
\citep{KittelKroemer1980,{TaylorSimsBarreraAllanMackrodt1999}}.
Thus, the governing equations are
\begin{equation}
    \frac{\partial \mathcal{F}}{\partial \mathbf{X}^i}
    =\frac{\partial \mathcal{E}}{\partial \mathbf{X}^i}+\frac{\hbar}{2}\sum_{r=1}^{3N} \frac{\partial \omega_r}{\partial \mathbf{X}^i}
    +\hbar\sum_{r=1}^{3N}\frac{1}{\exp \left(\frac{\hbar\omega_r}{k_BT}\right)-1}\frac{\partial \omega_r}{\partial \mathbf{X}^i}=\mathbf{0}.
\end{equation}

To compute the derivatives of the eigenvalues, we use the method
developed by Kantorovich \citep{Kantorovich1995}. Consider the
expansion of the elements of the dynamaical matrix
$\mathbf{\Phi}=\left[\Phi_{\alpha\beta}\right]$ about a configuration $\mathcal{B}$:
\begin{equation}
    \Phi_{\alpha\beta}\left(\left\{\mathbf{x}^i\right\}_{i\in
    \mathcal{L}}\right)=\Phi_{\alpha\beta}\left(\left\{\mathbf{X}^i\right\}_{i\in
    \mathcal{L}}\right)+ \sum_{i\in\mathcal{L}}\frac{\partial \Phi_{\alpha\beta}}{\partial \mathbf{X}^i}
    \left(\mathcal{B}\right)\cdot(\mathbf{x}^i-\mathbf{X}^i)+\cdots~~~~~\alpha,\beta=1,\ldots,3N.
\end{equation}
If the eigenvectors of $\mathbf{\Phi}$ are normalized to unity, the
perturbation expansion of eigenvalues would be
\citep{Kantorovich1995}
\begin{equation}
  \lambda_r\left(\left\{\mathbf{x}^i\right\}_{i\in
  \mathcal{L}}\right)=\lambda_r\left(\left\{\mathbf{X}^i\right\}_{i\in
  \mathcal{L}}\right)+
  \sum_{i\in\mathcal{L}}\sum_{\alpha , \beta=1}^{3N}U^{*}_{\alpha r}\frac{\partial \Phi_{\alpha \beta}} {\partial \mathbf{X}^i}U_{\beta r}\cdot(\mathbf{x}^i-\mathbf{X}^i)+\cdots,
\end{equation}
where $*$ denotes conjugate transpose and
$\mathbf{U}=\left[U_{\alpha\beta}\right]$ is the matrix of eigenvectors of $\mathbf{\Phi}=[\Phi_{\alpha\beta}]$, which are
normalized to unity. Since higher order terms in the
above expansion contain $(\mathbf{x}^i-\mathbf{X}^i)^n$ with $n \in
\mathbb{N} \geq 2$, all of them vanish for calculating the first
derivatives of eigenvalues at $\mathbf{x}^i = \mathbf{X}^i$. Hence,  we
can write
\begin{equation}
  \frac{\partial \lambda_r}{\partial \mathbf{x}^i}\Big|_{\mathbf{x}^i=\mathbf{X}^i}=\frac{\partial \lambda_r}{\partial \mathbf{X}^i}=
  \sum_{\alpha , \beta=1}^{3N}U^{*}_{\alpha r}\frac{\partial \Phi_{\alpha \beta}} {\partial \mathbf{X}^i}U_{\beta r},
\end{equation}
and therefore
\begin{equation}
  \frac{\partial \omega_r}{\partial \mathbf{X}^i}=\frac{1}{2 m_{r}\omega_{r}}
  \sum_{\alpha , \beta=1}^{3N}U^{*}_{\alpha r}\frac{\partial \Phi_{\alpha \beta}} {\partial \mathbf{X}^i}U_{\beta r}.
\end{equation}

For minimizing the free energy, depending on the chosen numerical
method, one may need the second derivatives of the eigenvalues as
well. We can extend the above procedure and consider higher
order terms to obtain higher order derivatives. The numerical
method used in this paper for minimizing the free energy will be discussed in detail in the
sequel.

\subsection{Perfect Crystals} Let us reformulate the classical theory of lattice dynamics
\citep{BornHuang1998,Mar71,Dove1993} in our notation for a perfect
crystal. This will make the formulation for defective crystals
clearer. Let us assume that we are given a multi-lattice
$\mathcal{L}$ with $N$ simple sublattices, i.e. $\mathcal{L}=\bigsqcup_{I=1}^{N}\mathcal{L}_I$. Let us denote the equilibrium position of $i\in\mathcal{L}$ by
$\mathbf{X}^i$, i.e.
\begin{equation}
    \frac{\partial }{\partial \mathbf{x}^i}\Big|_{\mathbf{x}^i=\mathbf{X}^i}\mathcal{E}\left(\{\mathbf{x}^j\}_{j\in\mathcal{L}}\right)=\mathbf{0}~~~~~\forall i\in\mathcal{L}.
\end{equation}
Atoms of the multi-lattice move from this equilibrium
configuration due to thermal vibrations. Let us denote the dynamic
position of atom $i\in\mathcal{L}$ by
$\mathbf{x}^i=\mathbf{x}^i(t)$. We now look for a wave-like
solution of the following form for $i\in\mathcal{L}_I$
\begin{equation}
    \mathbf{u}^i:=\mathbf{x}^i-\mathbf{X}^i=\frac{1}{\sqrt{m_I}}\mathbf{U}^{I}(\mathbf{k})~e^{\mathbbm{i}\left(\mathbf{k}\cdot\mathbf{X}^i-\omega(\mathbf{k})t\right)},
\end{equation}
where $\mathbbm{i}=\sqrt{-1}$, $\omega(\mathbf{k})$ is the frequency
at wave number $\mathbf{k}\in\textsf{B}$, $\textsf{B}$ is the
first Brillouin zone of the sublattices, and $\mathbf{U}^I$ is the
polarization vector. Note that we are assuming that $m_I\neq
0$.\footnote{For shell potentials, for example, shells are massless
and one obtains an effective dynamical matrix for cores as will be
explained in the sequel.} Note also that the displacements $\mathbf{x}^i(t)$ are time dependent and are deviations from the average temperature-dependent configuration $\mathbf{X}^i=\mathbf{X}^i(T)$.

Hamiltonian of this system has the following form
\begin{equation}
    \mathcal{H}\left(\{\mathbf{x}^i\}_{i\in\mathcal{L}}\right)=\frac{1}{2}\sum_{i\in\mathcal{L}}m_i|\dot{\mathbf{x}}^i|^2
    +\mathcal{E}\left(\{\mathbf{x}^i\}_{i\in\mathcal{L}}\right).
\end{equation}
Because of translation invariance of energy, it would be enough to
look at the equations of motion for the unit cell
$\mathbf{0}\in\mathbb{Z}^3$. These read $m_I\ddot{\mathbf{x}}^I=-\frac{\partial \mathcal{E}}{\partial \mathbf{x}^I},~I=1,...,N$. Note that
\begin{equation}
    m_I\ddot{\mathbf{x}}^I=-\sqrt{m_I}\mathbf{U}^I(\mathbf{k})\omega(\mathbf{k})^2~e^{\mathbbm{i}\left(\mathbf{k}\cdot\mathbf{X}^I-\omega(\mathbf{k})t\right)}.
\end{equation}
The idea of harmonic lattice dynamics is to linearize the forcing
term, i.e., to look at the following
linearized equations of motion.
\begin{equation}
    m_I\ddot{\mathbf{x}}^I=-\sum_{j\in\mathcal{L}}\frac{\partial^2 \mathcal{E}}{\partial \mathbf{x}^j\partial \mathbf{x}^I}(\mathcal{B})\mathbf{u}^j
    =-\sum_{J=1}^{N}\sum_{j\in\mathcal{L}_J}\frac{\partial^2 \mathcal{E}}{\partial \mathbf{x}^j\partial \mathbf{x}^I}(\mathcal{B})\mathbf{u}^j~~~~~I=1,...,N.
\end{equation}
Note that for $j\in\mathcal{L}_J$
\begin{equation}
    \mathbf{u}^j=\frac{1}{\sqrt{m_J}}\mathbf{U}^{J}(\mathbf{k})~e^{\mathbbm{i}\left(\mathbf{k}\cdot\mathbf{X}^j-\omega(\mathbf{k})t\right)}.
\end{equation}
Therefore, equations of motion read
\begin{equation}
    \omega(\mathbf{k})^2\mathbf{U}^I(\mathbf{k})=\sum_{J=1}^{N}\mathbf{D}_{IJ}(\mathbf{k})\mathbf{U}^J(\mathbf{k}),
\end{equation}
where
\begin{equation}
    \mathbf{D}_{IJ}=\frac{1}{\sqrt{m_Im_J}}\sum_{j\in\mathcal{L}_J}e^{\mathbbm{i}\mathbf{k}\cdot(\mathbf{X}^j-\mathbf{X}^I)}\frac{\partial^2 \mathcal{E}}{\partial \mathbf{x}^j\partial
    \mathbf{x}^I}(\mathcal{B}),
\end{equation}
are the sub-dynamical matrices. The case $I=J$ should be treated
carefully. We know that as a result of translation invariance of
energy
\begin{equation}
    \frac{\partial^2 \mathcal{E}}{\partial \mathbf{x}^I\partial \mathbf{x}^I}(\mathcal{B})=-\sum_{\begin{subarray}{l} j\in\mathcal{L}\\j\neq I \end{subarray}}
    \frac{\partial^2 \mathcal{E}}{\partial \mathbf{x}^j\partial \mathbf{x}^I}(\mathcal{B}).
\end{equation}
Thus
\begin{equation}
    \mathbf{D}_{II}=\frac{1}{m_I}\sum_{\begin{subarray}{l} j\in\mathcal{L}_I\\j\neq I \end{subarray}}
    e^{\mathbbm{i}\mathbf{k}\cdot(\mathbf{X}^j-\mathbf{X}^I)}\frac{\partial^2 \mathcal{E}}{\partial \mathbf{x}^j\partial \mathbf{x}^I}(\mathcal{B})
    -\frac{1}{m_I}\sum_{\begin{subarray}{l} j\in\mathcal{L}\\j\neq I \end{subarray}}\frac{\partial^2 \mathcal{E}}{\partial \mathbf{x}^j\partial
    \mathbf{x}^I}(\mathcal{B}).
\end{equation}
Finally, the dynamical matrix of the bulk crystal is defined as
\begin{equation}
    \mathbf{D}(\mathbf{k})=
    \left(%
    \begin{array}{cccc}
      \mathbf{D}_{11}(\mathbf{k}) & \mathbf{D}_{12}(\mathbf{k}) & \ldots & \mathbf{D}_{1N}(\mathbf{k}) \\
      \mathbf{D}_{21}(\mathbf{k}) & \mathbf{D}_{22}(\mathbf{k}) & \ldots & \mathbf{D}_{2N}(\mathbf{k}) \\
      \vdots & \vdots & \ddots & \vdots \\
      \mathbf{D}_{N1}(\mathbf{k}) & \mathbf{D}_{N2}(\mathbf{k}) & \ldots & \mathbf{D}_{NN}(\mathbf{k}) \\
    \end{array}%
    \right)\in \mathbb{R}^{3N \times 3N}.
\end{equation}
Let us denote the $3N$ eigenvalues of $\mathbf{D}(\mathbf{k})$ by
$\lambda_i(\mathbf{k}),~i=1,...,3N$. It is a well-known fact that
the dynamical matrix is Hermitian and hence all its eigenvalues
$\lambda_i$ are real. The crystal is stable if and only if
$\lambda_i>0~~~\forall~i$.

Free energy of the unit cell is now written as
\begin{equation}
    \mathcal{F}\left(\{\mathbf{X}^j\}_{j\in\mathcal{L}},T\right)=\mathcal{E}\left(\{\mathbf{X}^j\}_{j\in\mathcal{L}}\right)+\sum_{\mathbf{k}}\sum_{i=1}^{3N}\frac{1}{2}\hbar
    \omega_i(\mathbf{k})
    +\sum_{\mathbf{k}}\sum_{i=1}^{3N}k_BT\ln\left[1-\exp\left(-\frac{\hbar
    \omega_i(\mathbf{k})}{k_BT}\right)\right],
\end{equation}
where $\omega_i=\sqrt{\lambda_i}$ \footnote{Note that this is consistent with Eq. (\ref{Frequencies}) as we are using mass-reduced displacements.} and a finite sum over \textbf{k}-points
is used to approximate the integral over the first Brillouin zone of
the phonon density of states. The second term on the right-hand side
is the zero-point energy and the last term is the vibrational
entropy. For the optimum configuration
$\left\{\mathbf{X}^j\right\}_{j\in\mathcal{L}}$ at temperature $T$,
we have
\begin{equation}
    \frac{\partial \mathcal{F}}{\partial \mathbf{X}^j}=\frac{\partial \mathcal{E}}{\partial\mathbf{X}^j}+
    \sum_{\mathbf{k}}\sum_{i=1}^{3N}\left\{ \frac{\hbar}{2\omega_i(\mathbf{k})}\left(
    \frac{1}{2}+\frac{1}{\exp\left(\frac{\hbar \omega_i(\mathbf{k})}{k_BT}\right)-1}\right)
    \frac{\partial \omega^2_i(\mathbf{k})}{\partial \mathbf{X}^j}\right\}
    =\mathbf{0},~~~~~~~j=1,...,N.
\end{equation}
Here using the same procedure as in the pervious section, one can
calculate the derivatives of the eigenvalues as follows
\begin{equation}\label{eig-der}
  \frac{\partial \omega^2_{i} \left(\mathbf{k}\right)}{\partial \mathbf{X}^j}=
  \sum_{\alpha , \beta=1}^{3N}U^{*}_{\alpha i}\left(\mathbf{k}\right)
  \frac{\partial D^{\alpha \beta}  \left(\mathbf{k}\right)} {\partial \mathbf{X}^j}U_{\beta i}\left(\mathbf{k}\right),
\end{equation}
where
$\mathbf{U}\left(\mathbf{k}\right)=\left[U_{\alpha\beta}\left(\mathbf{k}\right)\right]\in
\mathbb{R}^{3N \times 3N}$ is the matrix of the eigenvectors of
$\mathbf{D}(\mathbf{k})=\left[D^{\alpha\beta}\left(\mathbf{k}\right)\right]$,
which are normalized to unity.

\subsection{Lattices with Massless Particles}

Let us next consider a lattice in
which some particles are assumed to be massless. The best
well-known model with this property is the so-called ``shell
model" \citep{DickOverhauser1964}. Let us assume that the unit
cell has $N$ particles (ions), each composed of a core and a (massless) shell. The
lattice $\mathcal{L}$ is partitioned as
\begin{equation}
    \mathcal{L}=\mathcal{L}^c\bigsqcup\mathcal{L}^s=\bigsqcup_{I=1}^{N}\left(\mathcal{L}^{c}_I\bigsqcup\mathcal{L}^{s}_I\right).
\end{equation}
Position vectors of core and shell of ion $i$ are denoted by
$\mathbf{x}_c^i$ and $\mathbf{x}_s^i$, respectively. Given a
configuration $\left\{\mathbf{x}^i\right\}_{i\in \mathcal{L}}$,
equations of motion for the fundamental unit cell read
\begin{equation}
    m_I \ddot{\mathbf{x}}_c^I =-\frac{\partial\mathcal{E}}{\partial \mathbf{x}_c^I},~~~
  \mathbf{0} =-\frac{\partial\mathcal{E}}{\partial
  \mathbf{x}_s^I},~~~~~~~~~~I=1,...,N.
\end{equation}
Assuming that cores and shells are at a static equilibrium
configuration, equations of motion in the harmonic approximation
read
\begin{eqnarray}
  \label{shell-equilibrium1} m_I \ddot{\mathbf{u}}_c^I &=& -\sum_{J=1}^{N}\sum_{j\in \mathcal{L}^c_J}\frac{\partial^2\mathcal{E}}{\partial \mathbf{x}_c^j\partial \mathbf{x}_c^I}\cdot\mathbf{u}^j_c
  -\sum_{J=1}^{N}\sum_{j\in \mathcal{L}^s_J}\frac{\partial^2\mathcal{E}}{\partial \mathbf{x}_s^j\partial \mathbf{x}_c^I}\cdot\mathbf{u}^j_s,~~~~~~~~~~I=1,...,N, \\
  \label{shell-equilibrium2} \mathbf{0} &=& -\sum_{J=1}^{N}\sum_{j\in \mathcal{L}^c_J}\frac{\partial^2\mathcal{E}}{\partial \mathbf{x}_c^j\partial \mathbf{x}_s^I}\cdot\mathbf{u}^j_c
  -\sum_{J=1}^{N}\sum_{j\in \mathcal{L}^s_J}\frac{\partial^2\mathcal{E}}{\partial \mathbf{x}_s^j\partial \mathbf{x}_s^I}\cdot\mathbf{u}^j_s,~~~~~~~~~~I=1,...,N.
\end{eqnarray}
Note that for $j\in\mathcal{L}_J$ we can write
\begin{equation}
    \mathbf{u}^j_c=\frac{1}{\sqrt{m_J}}\mathbf{U}^{J}_c(\mathbf{k})~e^{\mathbbm{i}\left(\mathbf{k}\cdot\mathbf{X}^j_c-\omega(\mathbf{k})t\right)},~~~
  \mathbf{u}^j_s=\mathbf{U}^{J}_s(\mathbf{k})~e^{\mathbbm{i}\left(\mathbf{k}\cdot\mathbf{X}^j_s-\omega(\mathbf{k})t\right)}~~~~~~\mathbf{k}\in
  \textsf{B},
\end{equation}
where $\textsf{B}$ is the first Brillouin zone of
$\mathcal{L}_I^c$ (or $\mathcal{L}_I^s$). Thus,
(\ref{shell-equilibrium1}) and (\ref{shell-equilibrium2}) can be
simplified to read
\begin{eqnarray}
  \label{condensation}  && \sum_{J=1}^{N}\mathbf{D}_{IJ}^{cc}\mathbf{U}^J_c(\mathbf{k})+\sum_{J=1}^{N}\mathbf{D}_{IJ}^{cs}\mathbf{U}^J_s(\mathbf{k})=\omega^2(\mathbf{k})\mathbf{U}_c^I(\mathbf{k})~~~~~~~~I=1,...,N, \\
  \label{condensation1} && \sum_{J=1}^{N}\mathbf{D}_{IJ}^{sc}\mathbf{U}^J_c(\mathbf{k})+\sum_{J=1}^{N}\mathbf{D}_{IJ}^{ss}\mathbf{U}^J_s(\mathbf{k})=\mathbf{0}~~~~~~~~~~~~~~~~~~~~~~I=1,...,N,
\end{eqnarray}
where
\begin{eqnarray}
  && \mathbf{D}_{IJ}^{cc}=\frac{1}{\sqrt{m_Im_J}}\sum_{j\in \mathcal{L}^c_J}\frac{\partial^2\mathcal{E}}{\partial \mathbf{x}_c^j\partial
    \mathbf{x}_c^I}e^{\mathbbm{i}\mathbf{k}\cdot(\mathbf{X}^j_c-\mathbf{X}_c^I)},~~
    \mathbf{D}_{IJ}^{cs}=\frac{1}{\sqrt{m_I}}\sum_{j\in \mathcal{L}^s_J}\frac{\partial^2\mathcal{E}}{\partial \mathbf{x}_s^j\partial
    \mathbf{x}_c^I}e^{\mathbbm{i}\mathbf{k}\cdot(\mathbf{X}^j_s-\mathbf{X}^I_c)} \nonumber \\
  && \mathbf{D}_{IJ}^{sc}=\frac{1}{\sqrt{m_J}}\sum_{j\in \mathcal{L}^c_J}\frac{\partial^2\mathcal{E}}{\partial \mathbf{x}_c^j\partial
    \mathbf{x}_s^I}e^{\mathbbm{i}\mathbf{k}\cdot(\mathbf{X}^j_c-\mathbf{X}^I_s)},~~~~~~
    \mathbf{D}_{IJ}^{ss}=\sum_{j\in \mathcal{L}^s_J}\frac{\partial^2\mathcal{E}}{\partial \mathbf{x}_s^j\partial
    \mathbf{x}_s^I}e^{\mathbbm{i}\mathbf{k}\cdot(\mathbf{X}^j_s-\mathbf{X}^I_s)}.
\end{eqnarray}
Eqs. (\ref{condensation}) and (\ref{condensation1}) can be
rewritten as
\begin{equation}
    \mathbf{D}_{cc}\mathbf{U}_c+\mathbf{D}_{cs}\mathbf{U}_s=\omega^2\mathbf{U}_c~~~~~\textrm{and}~~~~~\mathbf{U}_s=-\mathbf{D}_{ss}^{-1}\mathbf{D}_{sc}\mathbf{U}_c,
\end{equation}
where
\begin{eqnarray}
  && \mathbf{U}_c=\left(%
\begin{array}{c}
  \mathbf{U}_c^1 \\
  \vdots \\
  \mathbf{U}_c^N \\
\end{array}%
\right),~\mathbf{U}_s=\left(%
\begin{array}{c}
  \mathbf{U}_s^1 \\
  \vdots \\
  \mathbf{U}_s^N \\
\end{array}%
\right), \\
  && \nonumber \\
  && \mathbf{D}_{cc}=\left(%
\begin{array}{ccc}
  \mathbf{D}_{11}^{cc} & \hdots & \mathbf{D}_{1N}^{cc} \\
  \vdots & \ddots & \vdots \\
  \mathbf{D}_{N1}^{cc} & \hdots & \mathbf{D}_{NN}^{cc} \\
\end{array}%
\right),~\mathbf{D}_{cs}=\left(%
\begin{array}{ccc}
  \mathbf{D}_{11}^{cs} & \hdots & \mathbf{D}_{1N}^{cs} \\
  \vdots & \ddots & \vdots \\
  \mathbf{D}_{N1}^{cs} & \hdots & \mathbf{D}_{NN}^{cs} \\
\end{array}%
\right),  \\
  && \nonumber \\
  && \mathbf{D}_{sc}=\left(%
\begin{array}{ccc}
  \mathbf{D}_{11}^{sc} & \hdots & \mathbf{D}_{1N}^{sc} \\
  \vdots & \ddots & \vdots \\
  \mathbf{D}_{N1}^{sc} & \hdots & \mathbf{D}_{NN}^{sc} \\
\end{array}%
\right),~\mathbf{D}_{ss}=\left(%
\begin{array}{ccc}
  \mathbf{D}_{11}^{ss} & \hdots & \mathbf{D}_{1N}^{ss} \\
  \vdots & \ddots & \vdots \\
  \mathbf{D}_{N1}^{ss} & \hdots & \mathbf{D}_{NN}^{ss} \\
\end{array}%
\right).
\end{eqnarray}
Finally, the effective dynamical problem for cores can be written
as
\begin{equation}\
    \mathbf{D}(\mathbf{k})\mathbf{U}_c(\mathbf{k})=\omega(\mathbf{k})^2\mathbf{U}_c(\mathbf{k}),
\end{equation}
where
\begin{equation}\
    \mathbf{D}(\mathbf{k})=\mathbf{D}_{cc}(\mathbf{k})-\mathbf{D}_{cs}(\mathbf{k})\mathbf{D}_{ss}^{-1}(\mathbf{k})\mathbf{D}_{sc}(\mathbf{k}),
\end{equation}
is the effective dynamical matrix. Note that $\mathbf{D}_{cs}$ and
$\mathbf{D}_{sc}$ are not Hermitian but
$\mathbf{D}_{cs}\mathbf{D}_{ss}^{-1}\mathbf{D}_{sc}$ is.

The diagonal submatrices of $\mathbf{D}$, i.e.
$\mathbf{D}_{II}^{cc}$ and $\mathbf{D}_{II}^{ss}$ should be
calculated considering the translation invariance of energy,
namely
\begin{eqnarray}
  && \mathbf{D}_{II}^{cc}=\frac{1}{m_I}\sum_{\begin{subarray}{l} j\in\mathcal{L}^c_I\\j\neq Ic \end{subarray}}\frac{\partial^2\mathcal{E}}{\partial \mathbf{x}_c^j\partial
    \mathbf{x}_c^I}e^{\mathbbm{i}\mathbf{k}\cdot(\mathbf{X}^j_c-\mathbf{X}_c^I)}
    -\frac{1}{m_I}\sum_{\begin{subarray}{l} j\in\mathcal{L}\\j\neq Ic \end{subarray}}\frac{\partial^2\mathcal{E}}{\partial \mathbf{x}^j\partial
    \mathbf{x}_c^I}, \\
  && \mathbf{D}_{II}^{ss}=\sum_{\begin{subarray}{l} j\in\mathcal{L}^s_I\\j\neq Is \end{subarray}}\frac{\partial^2\mathcal{E}}{\partial \mathbf{x}_s^j\partial
    \mathbf{x}_s^I}e^{\mathbbm{i}\mathbf{k}\cdot(\mathbf{X}^j_s-\mathbf{X}^I_s))}
    -\sum_{\begin{subarray}{l} j\in\mathcal{L}\\j\neq Is \end{subarray}}\frac{\partial^2\mathcal{E}}{\partial \mathbf{x}^j\partial
    \mathbf{x}_s^I}.
\end{eqnarray}
Denoting the $3N$ eigenvalues of $\mathbf{D}(\mathbf{k})$ by
$\lambda_i(\mathbf{k})= \omega_i^2(\mathbf{k})$, free energy of
the unit cell is expressed as
\begin{equation}
    \mathcal{F}\left(\{\mathbf{X}^j_c,\mathbf{X}^j_s\}_{j\in\mathcal{L}},T\right)=\mathcal{E}\left(\{\mathbf{X}^j_c,\mathbf{X}^j_s\}_{j\in\mathcal{L}}\right)
    +\sum_{\mathbf{k}}\sum_{i=1}^{3N}\left\{ \frac{1}{2}\hbar  \omega_i(\mathbf{k})
    +k_BT\ln\left[1-\exp\left(\!-\frac{\hbar \omega_i(\mathbf{k})}{k_BT}\right)\right]\right\}.
\end{equation}
Therefore, for the optimum configuration
$\left\{\mathbf{X}^j_c,\mathbf{X}^j_s\right\}_{j\in\mathcal{L}}$ at
temperature $T$ we have
\begin{eqnarray}
  && \frac{\partial \mathcal{F}}{\partial \mathbf{X}^j_c}=\frac{\partial \mathcal{E}}{\partial \mathbf{X}^j_c}+
  \sum_{\mathbf{k}}\sum_{i=1}^{3N}\left\{\frac{\hbar}{2\omega_i(\mathbf{k})}\left(
    \frac{1}{2}+\frac{1}{\exp\left(\frac{\hbar \omega_i(\mathbf{k})}{k_BT}\right)-1}\right)
    \frac{\partial \omega^2_i(\mathbf{k})}{\partial \mathbf{X}^j_c}\right\}
    =\mathbf{0}, \\
  && \frac{\partial \mathcal{F}}{\partial \mathbf{X}^j_s}=\frac{\partial \mathcal{E}}{\partial \mathbf{X}^j_s}+
  \sum_{\mathbf{k}}\sum_{i=1}^{3N}\left\{\frac{\hbar}{2\omega_i(\mathbf{k})}\left(
    \frac{1}{2}+\frac{1}{\exp\left(\frac{\hbar \omega_i(\mathbf{k})}{k_BT}\right)-1}\right)
    \frac{\partial \omega^2_i(\mathbf{k})}{\partial \mathbf{X}^j_s}\right\}=\mathbf{0},
\end{eqnarray}
where the derivatives of eigenvalues are given by
\begin{equation}
  \frac{\partial \omega^2_{i} \left(\mathbf{k}\right)}{\partial \mathbf{X}^j_c}=
  \sum_{\alpha , \beta=1}^{3N}V^{*}_{\alpha i}\left(\mathbf{k}\right)
  \frac{\partial D_{\alpha \beta}  \left(\mathbf{k}\right)} {\partial \mathbf{X}^j_c}V_{\beta i}\left(\mathbf{k}\right),
  ~~~
  \frac{\partial \omega^2_{i} \left(\mathbf{k}\right)}{\partial \mathbf{X}^j_s}=
  \sum_{\alpha , \beta=1}^{3N}V^{*}_{\alpha i}\left(\mathbf{k}\right)
  \frac{\partial D_{\alpha \beta}  \left(\mathbf{k}\right)} {\partial \mathbf{X}^j_s}V_{\beta i}\left(\mathbf{k}\right),
\end{equation}
where
$\mathbf{V}\left(\mathbf{k}\right)=\left[V_{\alpha\beta}\left(\mathbf{k}\right)\right]\in
\mathbb{R}^{3N \times 3N}$ is the matrix of the eigenvectors of
$\mathbf{D}(\mathbf{k})=\left[D_{\alpha\beta}\left(\mathbf{k}\right)\right]$, which are normalized to unity.

\subsection{Defective Crystals}
 Without loss of generality, let us
consider a defective crystal with a 1-D symmetry reduction
\citep{Yavari2005a}, i.e.
\begin{equation}\label{1Dpartition}
    \mathcal{L}=\bigsqcup_{J=1}^{N}\bigsqcup_{\beta\in\mathbb{Z}}\mathcal{L}_{J\beta}.
\end{equation}
Note that $j=J\beta$ means that the atom $j$ is in the
$\beta$\emph{th} equivalence class of the $J$\emph{th} sublattice.
For this atom the thermal displacement vector is assumed to have
the following form
\begin{equation}
    \mathbf{u}^j=\frac{1}{\sqrt{m_J}}\mathbf{U}^{J\beta}(\mathbf{k})~e^{\mathbbm{i}\left(\mathbf{k}\cdot\mathbf{X}^j-\omega(\mathbf{k})t\right)},~~~\mathbf{k}\in
    \textsf{B},
\end{equation}
where $\textsf{B}$ is the first Brillouin zone of $\mathcal{L}_J$.
Equations of motion in this case read
\begin{equation}
    \omega(\mathbf{k})^2\mathbf{U}^{I\alpha}(\mathbf{k})=\sum_{J=1}^{N}\sum_{\beta\in\mathbb{Z}}\mathbf{D}_{I\alpha
    J\beta}(\mathbf{k})\mathbf{U}^{J\beta}(\mathbf{k}),
\end{equation}
where
\begin{equation}
    \mathbf{D}_{I\alpha J\beta}=\frac{1}{\sqrt{m_Im_J}}\sum_{j\in\mathcal{L}_{J\beta}}
    e^{\mathbbm{i}\mathbf{k}\cdot(\mathbf{X}^{j}-\mathbf{X}^{I\alpha})}\frac{\partial^2 \mathcal{E}}{\partial \mathbf{x}^{I\alpha}\partial
    \mathbf{x}^j}(\mathcal{B}),
\end{equation}
are the dynamical sub-matrices. The sub-matrices
$\mathbf{D}_{I\alpha I\alpha}$ have the following simplified form
\begin{equation}
    \mathbf{D}_{I\alpha I\alpha}=\frac{1}{m_I}\sum_{j\in\mathcal{L}_{I\alpha}}
    e^{\mathbbm{i}\mathbf{k}\cdot(\mathbf{X}^{j}-\mathbf{X}^{I\alpha})}\frac{\partial^2 \mathcal{E}}{\partial \mathbf{x}^{I\alpha}\partial
    \mathbf{x}^j}(\mathcal{B}).
\end{equation}
Note that
\begin{equation}
    \frac{\partial^2 \mathcal{E}}{\partial \mathbf{x}^{I\alpha}\partial \mathbf{x}^{I\alpha}}(\mathcal{B})
    =-\sum_{\begin{subarray}{l} j\in\mathcal{L}\\j\neq I\alpha \end{subarray}}\frac{\partial^2 \mathcal{E}}{\partial \mathbf{x}^{I\alpha}\partial
    \mathbf{x}^j}(\mathcal{B}).
\end{equation}
Thus
\begin{equation}
    \mathbf{D}_{I\alpha I\alpha}=\frac{1}{m_I}\sum_{\begin{subarray}{l} j\in\mathcal{L}_{I\alpha}\\j\neq I\alpha \end{subarray}}
    e^{\mathbbm{i}\mathbf{k}\cdot(\mathbf{X}^{j}-\mathbf{X}^{I\alpha})}\frac{\partial^2 \mathcal{E}}{\partial \mathbf{x}^{I\alpha}\partial \mathbf{x}^j}(\mathcal{B})
    -\frac{1}{m_I}\sum_{\begin{subarray}{l} j\in\mathcal{L}\\j\neq I\alpha \end{subarray}}\frac{\partial^2 \mathcal{E}}{\partial \mathbf{x}^{I\alpha}\partial
    \mathbf{x}^j}(\mathcal{B}).
\end{equation}
It is seen that for a defective crystal the dynamical matrix is
infinite dimensional.

As an approximation, similar to that presented in
\citep{LesarNajafabadiSrolovitz1989} as the local quasiharmonic
approximation, one can assume that given a unit cell, only a finite
number of neighboring equivalence classes interact with its thermal
vibrations. One way of approximating the free energy would then be
to consider vibrational effects in a finite region around the defect
and study the convergence of the results as a function of the size of
the finite region. For similar ideas see
\citep{KesavasamyKrishnamurthy1978,KesavasamyKrishnamurthy1979}, and
\citep{FernandezMontiPasianot2000}. Here, we consider a finite
number of equivalence classes, say $-C\leq\alpha\leq C$, around the
defect and assume the temperature-dependent bulk configuration outside this region. As another
approximation we assume that only a finite number of equivalence
classes interact with a given equivalence class in calculating the
dynamical matrix, i.e. we write
\begin{equation}
    \mathcal{L}_i=\bigsqcup_{\alpha=-m}^{m}\bigsqcup_{I=1}^{N}\mathcal{L}_{I\alpha},
\end{equation}
where $\mathcal{L}_i$ is the neighboring set of atom $i$.
Therefore, the linearized equations of motion read
\begin{equation}
    \omega(\mathbf{k})^2\mathbf{U}^{I\alpha}(\mathbf{k})=\sum_{\beta=-m}^{m}\sum_{J=1}^{N}\mathbf{D}_{I\alpha
    J\beta}(\mathbf{k})\mathbf{U}^{J\beta}(\mathbf{k})~~~~~~~~\alpha=-C,...,C.
\end{equation}
Defining
\begin{equation}\
    \mathbf{U}_{\alpha}=\left(%
\begin{array}{c}
  \mathbf{U}^{1\alpha} \\
  \vdots \\
  \mathbf{U}^{N\alpha} \\
\end{array}%
\right)\in\mathbb{R}^{3N},
\end{equation}
we can write the equations of motion as follows
\begin{equation}
    \omega(\mathbf{k})^2\mathbf{U}_{\alpha}(\mathbf{k}) =
    \sum_{\beta=-m}^{m}\mathbf{A}_{\alpha \left(\alpha+\beta\right)}(\mathbf{k})\mathbf{U}_{\left(\alpha+\beta\right)}(\mathbf{k}),
\end{equation}
where
\begin{eqnarray}
  && \mathbf{A}_{\alpha \beta}=\left(%
\begin{array}{ccc}
  \mathbf{D}_{1\alpha1\beta} & \hdots & \mathbf{D}_{1\alpha N\beta} \\
  \vdots & \ddots & \vdots \\
  \mathbf{D}_{N\alpha1\beta} & \hdots & \mathbf{D}_{N\alpha N\beta} \\
\end{array}%
\right)\in\mathbb{R}^{3N\times 3N}.
\end{eqnarray}
Now considering the finite classes around the defect, we can write
the global equations of motion for the finite system as
\begin{equation}
    \mathbf{D}(\mathbf{k})\mathbf{U}(\mathbf{k})=\omega(\mathbf{k})^2\mathbf{U}(\mathbf{k}),
\end{equation}
where
\begin{equation}\
    \mathbf{U}(\mathbf{k})=\left(%
\begin{array}{c}
  \mathbf{U}_{-C} \\
  \vdots \\
  \mathbf{U}_{C} \\
\end{array}%
\right)\in\mathbb{R}^{M}, ~\mathbf{D}(\mathbf{k})=\left(%
\begin{array}{ccc}
  \mathbb{D}_{\left(-C\right)\left(-C\right)} & \hdots & \mathbb{D}_{\left(-C\right)C} \\
  \vdots & \ddots & \vdots \\
  \mathbb{D}_{C\left(-C\right)} & \hdots & \mathbb{D}_{CC} \\
\end{array}%
\right)\in\mathbb{R}^{M\times M},~~~M=3N\times(2C+1),
\end{equation}
and
\begin{equation}\
    \mathbb{D}_{\alpha\beta}= \left\{ %
\begin{array}{c}
  \mathbf{A}_{\alpha\beta}~~~~~~~~~|\alpha-\beta|\leq m, \\
  \\
  \mathbf{0}_{3N\times3N} ~~~~|\alpha-\beta|> m.\\
\end{array}%
\right.
\end{equation}
It is easy to show that $\mathbf{A}_{\alpha\beta}(\mathbf{k})=\mathbf{A}^{\ast}_{\beta\alpha}(\mathbf{k})$, i.e. the dynamical matrix $\mathbf{D}(\mathbf{k})$ is
Hermitian, and therefore has $M$ real eigenvalues. Note that the
defective crystal is stable if and only if
$\omega_i^2>0~~~\forall~i$.

Now we can write the free energy of the defective crystal as
\begin{equation}
    \mathcal{F}\left(\{\mathbf{X}^j\}_{j\in\mathcal{L}},T\right)=\mathcal{E}\left(\{\mathbf{X}^j\}_{j\in\mathcal{L}}\right)+\sum_{\mathbf{k}}\sum_{i=1}^{M}\left\{\frac{1}{2}\hbar
    \omega_i(\mathbf{k})
    +k_BT\ln\left[1-\exp\left(-\frac{\hbar
    \omega_i(\mathbf{k})}{k_BT}\right)\right]\right\}.
\end{equation}
In the optimum configuration
$\left\{\mathbf{X}^j\right\}_{j\in\mathcal{L}}$ at a finite
temperature $T$, we have
\begin{equation}
    \frac{\partial \mathcal{F}}{\partial \mathbf{X}^j}=\frac{\partial \mathcal{E}}{\partial
    \mathbf{X}^j}+
    \sum_{\mathbf{k}}\sum_{i=1}^{M}\left\{\frac{\hbar}{2\omega_i(\mathbf{k})}\left(
    \frac{1}{2}+\frac{1}{\exp\left(\frac{\hbar \omega_i(\mathbf{k})}{k_BT}\right)-1}\right)
    \frac{\partial \omega^2_i(\mathbf{k})}{\partial \mathbf{X}^j}\right\}=\mathbf{0},
\end{equation}
where the derivatives of the eigenvalues are calculated as follows
\begin{equation}\label{eig-der}
  \frac{\partial \omega^2_{i} \left(\mathbf{k}\right)}{\partial \mathbf{X}^j}=
  \sum_{\alpha , \beta=1}^{M}U^{*}_{\alpha i}\left(\mathbf{k}\right)
  \frac{\partial D_{\alpha \beta}  \left(\mathbf{k}\right)} {\partial \mathbf{X}^j}U_{\beta i}\left(\mathbf{k}\right),
\end{equation}
where
$\mathbf{U}\left(\mathbf{k}\right)=\left[U_{\alpha\beta}\left(\mathbf{k}\right)\right]\in
\mathbb{R}^{M \times M}$ is the matrix of the eigenvectors of
$\mathbf{D}(\mathbf{k})=\left[D_{\alpha\beta}\left(\mathbf{k}\right)\right]$,
which are normalized to unity.

\subsection{Defect Structure at Finite Temperatures} In the static
case, given a configuration
$\mathcal{B}'_0=\left\{{\mathbf{x}'}_0^i\right\}_{i\in\mathcal{L}}$,
one can calculate the energy and hence forces exactly, as the
potential energy is calculated by some given empirical interatomic
potentials. Suppose one starts with a reference configuration and
solves for the following harmonic problem:
\begin{equation}
    \sum_{j\in\mathcal{L}}\frac{\partial^2\mathcal{E}}{\partial\mathbf{x}^i\partial\mathbf{x}^j}\left(\mathcal{B}'_0\right)\cdot(\mathbf{x}^j-{\mathbf{x}'}^j_0)
    =-\frac{\partial\mathcal{E}}{\partial\mathbf{x}^i}\left(\mathcal{B}'_0\right)~~~~~~~\forall~i\in\mathcal{L}.
\end{equation}
This reference configuration could be some nominal (unrelaxed)
configuration. Then one can modify the reference configuration and
by modified Newton-Raphson iterations converge to an
equilibrium configuration
$\mathcal{B}_0=\left\{\mathbf{x}_0^i\right\}_{i\in\mathcal{L}}$
assuming that such a configuration exists \citep{Yavari2005a}. In
this configuration $\frac{\partial\mathcal{E}}{\partial\mathbf{x}^i}\left(\mathcal{B}_0\right)=\mathbf{0},~\forall~i\in\mathcal{L}$. $\mathcal{B}_0$ is now the starting configuration for lattice
dynamics.\footnote{If temperature is ``large", one can start with
equilibrium configuration of a lower temperature. This is what we do in our numerical examples as will be discussed in the sequel.} For a
temperature $T$, the defective crystal is in thermal equilibrium
if the free energy is minimized, i.e., if
\begin{equation}\label{entropy}
    \frac{\partial\mathcal{F}}{\partial\mathbf{X}^i}\left(\mathcal{B}\right)=\mathbf{0},~\forall~i\in\mathcal{L}.
\end{equation}
Solving this problem one can modify the reference configuration
and calculate the optimum configuration. This iteration would give
a configuration that minimizes the harmonically calculated free
energy. The next step then would be to correct for anharmonic
effects in the vibrational frequencies. One way of doing this
is to iteratively calculate the vibrational unbalanced
forces using higher order terms in the Taylor expansion.

There are many different optimization techniques to solve the
unconstrained minimization problem (\ref{entropy}). Here we only
consider two main methods that are usually more
efficient, namely those that require only the gradient and those
that require the gradient and the Hessian \citep{PressTVF1989}. In
problems in which the Hessian is available, the Newton method is
usually the most powerful. It is based on the following quadratic
approximation near the current configuration
\begin{equation}
    \mathcal{F}\left(\mathcal{B}^{k}+\tilde{\boldsymbol{\delta}}^k\right)=\mathcal{F}\left(\mathcal{B}^{k}\right)
    +\boldsymbol{\nabla}\mathcal{F}\left(\mathcal{B}^{k}\right)\cdot\tilde{\boldsymbol{\delta}}^k
    +\frac{1}{2}(\tilde{\boldsymbol{\delta}}^k)^{\textsf{T}}\cdot\mathbf{H}\left(\mathcal{B}^{k}\right)\cdot\tilde{\boldsymbol{\delta}}^k+o\left(|\tilde{\boldsymbol{\delta}}^k|^2\right),
\end{equation}
where $\tilde{\boldsymbol{\delta}}^k=\mathcal{B}^{k+1}-\mathcal{B}^{k}$. Now if
we differentiate the above formula with respect to
$\tilde{\boldsymbol{\delta}}^k$, we obtain Newton method for determining the
next configuration
$\mathcal{B}^{k+1}=\mathcal{B}^{k}+\tilde{\boldsymbol{\delta}}^k:~~\tilde{\delta}^k=-\mathbf{H}^{-1}\left(\mathcal{B}^{k}\right)\cdot\boldsymbol{\nabla}\mathcal{F}\left(\mathcal{B}^{k}\right)$. Here in order to converge to a local minimum the Hessian must be
positive definite.

One can use a perturbation method to obtain the second derivatives of the free energy but
as the dimension of a defective crystal increases, calculation of
these higher order derivatives may become numerically inefficient
\citep{TaylorBarreraAllanBarron1997} and so one may prefer to use
those methods that do not require the second derivatives. One such
method is the quasi-Newton method. The main idea behind this
method is to start from a positive-definite approximation to the
inverse Hessian and to modify this approximation in each iteration
using the gradient vector of that step. Close to the local
minimum, the approximate inverse Hessian approaches the true
inverse Hessian and we would have the quadratic convergence of
Newton method \citep{PressTVF1989}. There are different algorithms
for generating the approximate inverse Hessian. One of the most
well known is the Broyden-Fletcher-Goldfarb-Shanno (BFGS)
algorithm \citep{PressTVF1989}:
\begin{equation}\label{BFGS}
    \mathbf{A}^{i+1}=\mathbf{A}^i + \frac{\tilde{\boldsymbol{\delta}}^k\otimes\tilde{\boldsymbol{\delta}}^k}{(\tilde{\boldsymbol{\delta}}^k)^{\textsf{T}}\cdot\mathbf{\Delta}}
    -\frac{\left(\mathbf{A}^{i}\cdot\mathbf{\Delta}\right)\otimes\left(\mathbf{A}^{i}\cdot\mathbf{\Delta}\right)}
          {\mathbf{\Delta}^{\textsf{T}}\cdot\mathbf{A}^{i}\cdot\mathbf{\Delta}}
    +\left(\mathbf{\Delta}^{\textsf{T}}\cdot\mathbf{A}^{i}\cdot\mathbf{\Delta}\right)\mathbf{u}\otimes\mathbf{u},
\end{equation}
where $\mathbf{A}^{i}=\left(\mathbf{H}^i\right)^{-1}$, $\mathbf{\Delta}=\boldsymbol{\nabla}\mathcal{F}^{i+1}-\boldsymbol{\nabla}\mathcal{F}^{i}$, and
\begin{eqnarray}
  \mathbf{u}=\frac{\tilde{\boldsymbol{\delta}}^k}{(\tilde{\boldsymbol{\delta}}^k)^{\textsf{T}}\cdot\mathbf{\Delta}}-
  \frac{\mathbf{A}^{i}\cdot\mathbf{\Delta}}{\mathbf{\Delta}^{\textsf{T}}\cdot\mathbf{A}^{i}\cdot\mathbf{\Delta}}.
\end{eqnarray}
Calculating $\mathbf{A}^{i+1}$, one then should use
$\mathbf{A}^{i+1}$ instead of $\mathbf{H}^{-1}$ to update the current configuration for the next
configuration
$\mathcal{B}^{k+1}=\mathcal{B}^{k}+\tilde{\boldsymbol{\delta}}^k$.
If $\mathbf{A}^{i+1}$ is a poor approximation, then one may need to
perform a linear search to refine $\mathcal{B}^{k+1}$ before
starting the next iteration \citep{PressTVF1989}. As Taylor, et al.
\citep{TaylorBarreraAllanBarron1997} mention, since the dynamical
contributions to the Hessian are usually small, one can use only the
static part of the free energy $ \mathcal{E}$ to generate the first
approximation to the Hessian of the free energy. Therefore, we
propose the following quasiharmonic lattice dynamics algorithm based
on the quasi-Newton method:

 \vskip -0.7 in
\begin{picture}(150,150)
\framebox{\begin{minipage}[c]{4in}
\begin{itemize}
    \item [] Input data: $\mathcal{B}_0$ (or $\mathcal{B}_{T-\Delta T}$ for large $T$), $T$
    \item [$\vartriangleright$] Initialization
    \begin{itemize}
        \item [$\vartriangleright$] $\mathbf{H}^1=\mathbf{H}_{static}|_{\mathcal{B}=\mathcal{B}_0}$
    \end{itemize}
    \item [$\vartriangleright$] Do until convergence is achieved
    \begin{itemize}
        \item [$\vartriangleright$] $\mathbf{D}^{k}=\mathbf{D}(\mathcal{B}^k)$
        \item [$\vartriangleright$] Calculate $\nabla\mathcal{F}^{k}$
        \item [$\vartriangleright$] Use $\mathbf{A}^{k}$ to obtain $\mathcal{B}^{k+1}$
    \end{itemize}
    \item [$\vartriangleright$] End Do
    \item [$\vartriangleright$] End
\end{itemize}
\end{minipage}}
\end{picture}
\vskip 1.5 in

\section {\textbf{Lattice Dynamic Analysis of a Defective Lattice of Point Dipoles}}

In this section we consider a two-dimensional defective lattice of
dipoles. \citet{Westhaus1981} derived the normal mode
frequencies for a 2-D rectangular lattice of point dipoles using the
assumption that interacting dipoles have fixed length polarization vectors
that can only rotate around fixed lattice
sites. In this section, we relax these assumptions and in the next section will obtain the temperature-dependent structures of two $180^{\circ}$ domain walls.

Consider a defective lattice of dipoles in which each lattice point
represents a unit cell and the corresponding dipole is a measure of
the distortion of the unit cell with respect to a high symmetry
phase. Total energy of the lattice is assumed to have the following
three parts \citep{Yavari2005a}
\begin{equation}
    \mathcal{E}\left(\{\mathbf{x}^i,\mathbf{P}^i\}_{i\in\mathcal{L}}\right)=
    \mathcal{E}^{\textrm{d}}\left(\{\mathbf{x}^i,\mathbf{P}^i\}_{i\in\mathcal{L}}\right)
    +\mathcal{E}^{\textrm{short}}\left(\{\mathbf{x}^i\}_{i\in\mathcal{L}}\right)+\mathcal{E}^{\textrm{a}}\left(\{\mathbf{P}^i\}_{i\in\mathcal{L}}\right),
\end{equation}
where, $\mathcal{E}^{\textrm{d}}$, $\mathcal{E}^{\textrm{short}}$
and $\mathcal{E}^{\textrm{a}}$ are the dipole energy, short-range
energy, and anisotropy energy, respectively. The dipole energy has
the following form
\begin{equation}
    \mathcal{E}^{\textrm{d}}=\frac{1}{2}\sum_{\begin{subarray}{l} i,j\in \mathcal{L}\\~j\neq i \end{subarray}}
    \left\{\frac{\mathbf{P}^i\cdot\mathbf{P}^j}{|\mathbf{x}^i-\mathbf{x}^j|^3}
    -\frac{3\mathbf{P}^i\cdot(\mathbf{x}^i-\mathbf{x}^j)~\mathbf{P}^j\cdot(\mathbf{x}^i-\mathbf{x}^j)}
    {|\mathbf{x}^i-\mathbf{x}^j|^5}\right\}+\sum_{i\in\mathcal{L}}\frac{1}{2\alpha_i}\mathbf{P}^i\cdot\mathbf{P}^i,
\end{equation}
where $\alpha_i$ is the electric polarizability and is assumed to
be a constant for each sublattice. For the sake of simplicity, we assume that polarizability is temperature independent. The short-range energy is
modeled by a Lennard-Jones potential with the following form
\begin{equation}
    \mathcal{E}^{\textrm{short}}=\frac{1}{2}\sum_{\begin{subarray}{l} i,j\in
    \mathcal{L}\\~j\neq i \end{subarray}}
    4\epsilon_{ij}\left[\left(\frac{a_{ij}}{|\mathbf{x}^i-\mathbf{x}^j|}\right)^{12}-\left(\frac{a_{ij}}{|\mathbf{x}^i-\mathbf{x}^j|}\right)^6\right],
\end{equation}
where for a multi-lattice with two sublattices $a_{ij}$ and
$\epsilon_{ij}$ take values in the sets $\{a_{11},a_{12},a_{22}\}$
and $\{\epsilon_{11},\epsilon_{12},\epsilon_{22}\}$, respectively.
The anisotropy energy quantifies the tendency of the lattice to
remain in some energy wells and is assumed to have the following
form
\begin{equation}
    \mathcal{E}^{\textrm{a}}=\sum_{i\in\mathcal{L}}K_{A}|\mathbf{P}^i-\mathbf{P}_1|^2~|\mathbf{P}^i-\mathbf{P}_2|^2.
\end{equation}
This means that the dipoles prefer to have values in the set
$\{\mathbf{P}_1,\mathbf{P}_2\}$.

Let
$\mathcal{S}=\left(\{\mathbf{X}^i,\mathbf{P}^i\}_{i\in\mathcal{L}}\right)$
be the equilibrium configuration (a local minimum of the energy),
i.e.
\begin{equation} \label{gov}
    \frac{\partial\mathcal{E}}{\partial \mathbf{X}^i}=\frac{\partial\mathcal{E}}{\partial
    \mathbf{P}^i}=\mathbf{0}~~~~~\forall~i \in
    \mathcal{L}.
\end{equation}
It was shown in \citep{Yavari2005a} how to find a static
equilibrium equation starting from a reference configuration. We
assume that this configuration is given and denote it by
$\mathcal{B}=\{\mathbf{X}^i,\mathbf{\bar{P}}^i\}_{i\in\mathcal{L}}$.
At a finite temperature $T$, ignoring the dipole inertia,
Hamiltonian of this system can be written as
\begin{equation}
    \mathcal{H}\left(\{\mathbf{x}^i,\mathbf{P}^i\}_{i\in\mathcal{L}}\right)=\frac{1}{2}\sum_{i\in\mathcal{L}}m_i|\dot{\mathbf{x}}^i|^2
    +\mathcal{E}\left(\{\mathbf{x}^i,\mathbf{P}^i\}_{i\in\mathcal{L}}\right).
\end{equation}
Equations of motion read
\begin{equation}\label{equations-motion}
   m_i \ddot{\mathbf{x}}^i = -\frac{\partial\mathcal{E}}{\partial \mathbf{x}^i},~~~\mathbf{0} = -\frac{\partial\mathcal{E}}{\partial
  \mathbf{P}^i}.
\end{equation}
Linearizing the equations of motion (\ref{equations-motion}) about the equilibrium configuration, we
obtain
\begin{eqnarray}
  -m_i\ddot{\mathbf{x}}^i &=& \frac{\partial^{2}\mathcal{E}}{\partial\mathbf{x}^i\partial\mathbf{x}^i}\left(\mathcal{B}\right)(\mathbf{x}^i-\mathbf{X}^i)
    +\sum_{j\in\mathcal{S}_i}\frac{\partial^{2}\mathcal{E}}{\partial\mathbf{x}^j\partial\mathbf{x}^i}\left(\mathcal{B}\right)(\mathbf{x}^j-\mathbf{X}^j)
    \nonumber \\
  && \label{equations-motion1} +\frac{\partial^{2}\mathcal{E}}{\partial\mathbf{P}^i\partial\mathbf{x}^i}\left(\mathcal{B}\right)(\mathbf{P}^i-\mathbf{\bar{P}}^i)
    +\sum_{j\in\mathcal{S}_i}\frac{\partial^{2}\mathcal{E}}{\partial\mathbf{P}^j\partial\mathbf{x}^i}\left(\mathcal{B}\right)(\mathbf{P}^j-\mathbf{\bar{P}}^j), \\
  \mathbf{0} &=& \frac{\partial^{2}\mathcal{E}}{\partial\mathbf{x}^i\partial\mathbf{P}^i}\left(\mathcal{B}\right)(\mathbf{x}^i-\mathbf{X}^i)+
    \sum_{j\in\mathcal{S}_i}\frac{\partial^{2}\mathcal{E}}{\partial\mathbf{x}^j\partial\mathbf{P}^i}\left(\mathcal{B}\right)(\mathbf{x}^j-\mathbf{X}^j) \nonumber\\
  && \label{equations-motion2}  +\frac{\partial^{2}\mathcal{E}}{\partial\mathbf{P}^i\partial\mathbf{P}^i}\left(\mathcal{B}\right)(\mathbf{P}^i-\mathbf{\bar{P}}^i)
    +\sum_{j\in\mathcal{S}_i}\frac{\partial^{2}\mathcal{E}}{\partial\mathbf{P}^j\partial\mathbf{P}^i}\left(\mathcal{B}\right)(\mathbf{P}^j-\mathbf{\bar{P}}^j),
\end{eqnarray}
where $\mathcal{S}_i=\mathcal{L}\setminus \{i\}$. Note that
\begin{eqnarray}
  &&
  \frac{\partial^{2}\mathcal{E}}{\partial\mathbf{P}^i\partial\mathbf{P}^i}\left(\mathcal{B}\right)=2K_A\left(|\mathbf{\bar{P}}^i-\mathbf{P}_1|^2+|\mathbf{\bar{P}}^i-\mathbf{P}_2|^2\right)
    \mathbf{I}+4K_A\left(\mathbf{\bar{P}}^i-\mathbf{P}_1\right)\otimes\left(\mathbf{\bar{P}}^i-\mathbf{P}_2\right) \nonumber\\
   && ~~~~~~~~~~~~~~~~~~~~+4K_A\left(\mathbf{\bar{P}}^i-\mathbf{P}_2\right)\otimes\left(\mathbf{\bar{P}}^i-\mathbf{P}_1\right)+\frac{1}{\alpha_i}\mathbf{I},
\end{eqnarray}
where $\mathbf{I}$ is the $2\times2$ identity matrix and $\otimes$
denotes tensor product.

For a defective crystal with a 1-D symmetry reduction the set
$\mathcal{L}$ can be partitioned as follows
\begin{equation}
    \mathcal{L}=\bigsqcup_{\alpha\in\mathbb{Z}}\bigsqcup_{I=1}^{N}\mathcal{L}_{I\alpha}.
\end{equation}
Let us define $\mathbf{u}^i=\mathbf{x}^i-\mathbf{X}^i,~~\mathbf{q}^i=\mathbf{P}^i-\mathbf{\bar{P}}^i$. Periodicity of the lattice allows us to write for
$i\in\mathcal{L}_{I\alpha}$
\begin{equation}
    \mathbf{u}^i=\frac{1}{\sqrt{m_I}}\mathbf{U}^{I\alpha}(\mathbf{k})~e^{\mathbbm{i}(\mathbf{k}\cdot\mathbf{X}^i-\omega(\mathbf{k})t)}
    ,~~\mathbf{q}^i=\mathbf{Q}^{I\alpha}(\mathbf{k})~e^{\mathbbm{i}(\mathbf{k}\cdot\mathbf{X}^i-\omega(\mathbf{k})t)},~~~\mathbf{k}\in \textsf{B}.
\end{equation}
Thus, Eq. (\ref{equations-motion1}) for $i=I\alpha$ can be
simplified to read
\begin{eqnarray}\label{eigen-value}
  && \omega(\mathbf{k})^2\mathbf{U}^{I\alpha}(\mathbf{k}) =
  \frac{1}{m_I}\frac{\partial^{2}\mathcal{E}}{\partial\mathbf{x}^{I\alpha}\partial\mathbf{x}^{I\alpha}}\left(\mathcal{B}\right)\mathbf{U}^{I\alpha}(\mathbf{k})
  +\sum_{J=1}^{N}\sum_{\beta\in\mathbb{Z}}~\sideset{}{'}{\sum}_{j\in \mathcal{L}_{J\beta}}\frac{1}{\sqrt{m_Im_J}}\frac{\partial^{2}\mathcal{E}}{\partial\mathbf{x}^j\partial\mathbf{x}^{I\alpha}}\left(\mathcal{B}\right)e^{\mathbbm{i}\mathbf{k}\cdot(\mathbf{X}^j-\mathbf{X}^{I\alpha})}
   \mathbf{U}^{J\beta}(\mathbf{k})
    \nonumber \\
  && ~~~~~+\frac{1}{\sqrt{m_I}}\frac{\partial^{2}\mathcal{E}}{\partial\mathbf{P}^{I\alpha}\partial\mathbf{x}^{I\alpha}}\left(\mathcal{B}\right)\mathbf{Q}^{I\alpha}(\mathbf{k})
    +\sum_{J=1}^{N}\sum_{\beta\in\mathbb{Z}}~\sideset{}{'}{\sum}_{j\in
    \mathcal{L}_{J\beta}}\frac{1}{\sqrt{m_I}}\frac{\partial^{2}\mathcal{E}}{\partial\mathbf{P}^j\partial\mathbf{x}^{I\alpha}}\left(\mathcal{B}\right)e^{\mathbbm{i}\mathbf{k}\cdot(\mathbf{X}^j-\mathbf{X}^{I\alpha})}
   \mathbf{Q}^{J\beta}(\mathbf{k}),
\end{eqnarray}
where a prime on summations means that the term corresponding to
$J\beta=I\alpha$ is excluded. Eq. (\ref{eigen-value}) can be
rewritten as
\begin{equation}\ \label{equations-motion3}
    \omega(\mathbf{k})^2\mathbf{U}^{I\alpha}(\mathbf{k}) = \sum_{J=1}^{N}\sum_{\beta\in\mathbb{Z}}\mathbf{D}^{xx}_{I\alpha J\beta}(\mathbf{k})\mathbf{U}^{J\beta}(\mathbf{k})
    +\sum_{J=1}^{N}\sum_{\beta\in\mathbb{Z}}\mathbf{D}^{xp}_{I\alpha J\beta}(\mathbf{k})\mathbf{Q}^{J\beta}(\mathbf{k}),
\end{equation}
where
\begin{eqnarray}
  && \mathbf{D}^{xx}_{I\alpha J\beta}(\mathbf{k}) = \delta_{\alpha \beta}\delta_{IJ}\frac{1}{m_I}\frac{\partial^{2}\mathcal{E}}{\partial\mathbf{x}^{I\alpha}\partial\mathbf{x}^{I\alpha}}\left(\mathcal{B}\right)
  +\sideset{}{'}{\sum}_{j\in \mathcal{L}_{J\beta}}\frac{1}{\sqrt{m_Im_J}}\frac{\partial^{2}\mathcal{E}}{\partial\mathbf{x}^j\partial\mathbf{x}^{I\alpha}}\left(\mathcal{B}\right)e^{\mathbbm{i}\mathbf{k}\cdot(\mathbf{X}^j-\mathbf{X}^{I\alpha})},
   \nonumber \\
  &&\mathbf{D}^{xp}_{I\alpha J\beta}(\mathbf{k}) = \delta_{\alpha \beta}\delta_{IJ}\frac{1}{\sqrt{m_I}}\frac{\partial^{2}\mathcal{E}}{\partial\mathbf{P}^{I\alpha}\partial\mathbf{x}^{I\alpha}}\left(\mathcal{B}\right)
  +\sideset{}{'}{\sum}_{j\in \mathcal{L}_{J\beta}}\frac{1}{\sqrt{m_I}}\frac{\partial^{2}\mathcal{E}}{\partial\mathbf{P}^j\partial\mathbf{x}^{I\alpha}}\left(\mathcal{B}\right)e^{\mathbbm{i}\mathbf{k}\cdot(\mathbf{X}^j-\mathbf{X}^{I\alpha})}.
\end{eqnarray}
Similarly, Eq. (\ref{equations-motion2}) can be simplified to read
\begin{eqnarray}
  && \frac{1}{\sqrt{m_I}}\frac{\partial^{2}\mathcal{E}}{\partial\mathbf{x}^{I\alpha}\partial\mathbf{P}^{I\alpha}}\left(\mathcal{B}\right)\mathbf{U}^{I\alpha}(\mathbf{k})
  +\sum_{J=1}^{N}\sum_{\beta\in\mathbb{Z}}~\sideset{}{'}{\sum}_{j\in \mathcal{L}_{J\beta}}\frac{1}{\sqrt{m_J}}\frac{\partial^{2}\mathcal{E}}{\partial\mathbf{x}^j\partial\mathbf{P}^{I\alpha}}\left(\mathcal{B}\right)e^{\mathbbm{i}\mathbf{k}\cdot(\mathbf{X}^j-\mathbf{X}^{I\alpha})}
   \mathbf{U}^{J\beta}(\mathbf{k})  \nonumber \\
  &&~~~+ \frac{\partial^{2}\mathcal{E}}{\partial\mathbf{P}^{I\alpha}\partial\mathbf{P}^{I\alpha}}\left(\mathcal{B}\right)\mathbf{Q}^{I\alpha}(\mathbf{k})
  +\sum_{J=1}^{N}\sum_{\beta\in\mathbb{Z}}~\sideset{}{'}{\sum}_{j\in \mathcal{L}_{J\beta}}\frac{\partial^{2}\mathcal{E}}{\partial\mathbf{P}^j\partial\mathbf{P}^{I\alpha}}\left(\mathcal{B}\right)e^{\mathbbm{i}\mathbf{k}\cdot(\mathbf{X}^j-\mathbf{X}^{I\alpha})}
   \mathbf{Q}^{J\beta}(\mathbf{k})=\mathbf{0}.
\end{eqnarray}
Or
\begin{equation}\ \label{equations-motion4}
    \sum_{J=1}^{N}\sum_{\beta\in\mathbb{Z}}\mathbf{D}^{px}_{I\alpha J\beta}(\mathbf{k})\mathbf{U}^{J\beta}(\mathbf{k})
    +\sum_{J=1}^{N}\sum_{\beta\in\mathbb{Z}}\mathbf{D}^{pp}_{I\alpha J\beta}(\mathbf{k})\mathbf{Q}^{J\beta}(\mathbf{k})=\mathbf{0},
\end{equation}
where
\begin{eqnarray}
  && \mathbf{D}^{px}_{I\alpha J\beta}(\mathbf{k}) = \delta_{\alpha \beta}\delta_{IJ}\frac{1}{\sqrt{m_I}}\frac{\partial^{2}\mathcal{E}}{\partial\mathbf{x}^{I\alpha}\partial\mathbf{P}^{I\alpha}}\left(\mathcal{B}\right)
  +\sideset{}{'}{\sum}_{j\in \mathcal{L}_{J\beta}}\frac{1}{\sqrt{m_J}}\frac{\partial^{2}\mathcal{E}}{\partial\mathbf{x}^j\partial\mathbf{P}^{I\alpha}}\left(\mathcal{B}\right)e^{\mathbbm{i}\mathbf{k}\cdot(\mathbf{X}^j-\mathbf{X}^{I\alpha})},
   \nonumber \\
  &&\mathbf{D}^{pp}_{I\alpha J\beta}(\mathbf{k}) = \delta_{\alpha \beta}\delta_{IJ}\frac{\partial^{2}\mathcal{E}}{\partial\mathbf{P}^{I\alpha}\partial\mathbf{P}^{I\alpha}}\left(\mathcal{B}\right)
  +\sideset{}{'}{\sum}_{j\in \mathcal{L}_{J\beta}} \frac{\partial^{2}\mathcal{E}}{\partial\mathbf{P}^j\partial\mathbf{P}^{I\alpha}}\left(\mathcal{B}\right)e^{\mathbbm{i}\mathbf{k}\cdot(\mathbf{X}^j-\mathbf{X}^{I\alpha})}.
\end{eqnarray}
We know that \citep{Yavari2005a}
\begin{equation} \label{invar1}
     \frac{\partial^{2}\mathcal{E}}{\partial\mathbf{x}^{I\alpha}\partial\mathbf{x}^{I\alpha}}\left(\mathcal{B}\right)=
    -\sideset{}{'}{\sum}_{j\in\mathcal{L}}\frac{\partial^{2}\mathcal{E}}{\partial\mathbf{x}^j\partial\mathbf{x}^{I\alpha}}\left(\mathcal{B}\right).
\end{equation}
And
\begin{equation} \label{invar2}
    \frac{\partial^{2}\mathcal{E}}{\partial\mathbf{x}^{I\alpha}\partial\mathbf{P}^{I\alpha}}\left(\mathcal{B}\right)
    =\frac{\partial^{2}\mathcal{E}}{\partial\mathbf{P}^{I\alpha}\partial\mathbf{x}^{I\alpha}}\left(\mathcal{B}\right)
    =-\sideset{}{'}{\sum}_{j\in\mathcal{L}}\frac{\partial^{2}\mathcal{E}}{\partial\mathbf{x}^j\partial\mathbf{P}^{I\alpha}}\left(\mathcal{B}\right).
\end{equation}
Before proceeding any further, let us first look at dynamical
matrix of the bulk lattice.

\paragraph{Dynamical Matrix for the Bulk Lattice.} In the case of
the bulk lattice we have
\begin{equation}
    \mathcal{L}=\bigsqcup_{I=1}^{N}\mathcal{L}_{I}.
\end{equation}
Periodicity of the lattice allows us to write for
$i\in\mathcal{L}_{I}$
\begin{equation}
    \mathbf{u}^i=\frac{1}{\sqrt{m_I}}\mathbf{U}^{I}(\mathbf{k})~e^{\mathbbm{i}(\mathbf{k}\cdot\mathbf{X}^i-\omega(\mathbf{k})t)}
    ,~~\mathbf{q}^i=\mathbf{Q}^{I}(\mathbf{k})~e^{\mathbbm{i}(\mathbf{k}\cdot\mathbf{X}^i-\omega(\mathbf{k})t)},~~~\mathbf{k}\in \textsf{B}.
\end{equation}
Thus, Eq. (\ref{equations-motion}) for $i=I$ is simplified to
read
\begin{eqnarray}
  && \omega(\mathbf{k})^2\mathbf{U}^{I}(\mathbf{k}) = \frac{1}{m_I}\frac{\partial^{2}\mathcal{E}}{\partial\mathbf{x}^{I}\partial\mathbf{x}^{I}}\left(\mathcal{B}\right)\mathbf{U}^{I}(\mathbf{k})
     +\sum_{J=1}^{N}~\sideset{}{'}{\sum}_{j\in \mathcal{L}_{J}}\frac{1}{\sqrt{m_Im_J}}\frac{\partial^{2}\mathcal{E}}{\partial\mathbf{x}^j\partial\mathbf{x}^{I}}\left(\mathcal{B}\right)e^{\mathbbm{i}\mathbf{k}\cdot(\mathbf{X}^j-\mathbf{X}^{I})}
   \mathbf{U}^{J}(\mathbf{k}) \nonumber \\
  && ~~~~~~~~~~~~~~~~~~~~+\frac{1}{\sqrt{m_I}}\frac{\partial^{2}\mathcal{E}}{\partial\mathbf{P}^{I}\partial\mathbf{x}^{I}}\left(\mathcal{B}\right)\mathbf{Q}^{I}(\mathbf{k})
  +\sum_{J=1}^{N}~\sideset{}{'}{\sum}_{j\in \mathcal{L}_{J}}\frac{1}{\sqrt{m_I}}\frac{\partial^{2}\mathcal{E}}{\partial\mathbf{P}^j\partial\mathbf{x}^{I}}\left(\mathcal{B}\right)e^{\mathbbm{i}\mathbf{k}\cdot(\mathbf{X}^j-\mathbf{X}^{I})}
   \mathbf{Q}^{J}(\mathbf{k}).
\end{eqnarray}
This can be rewritten as
\begin{equation}\
    \omega(\mathbf{k})^2\mathbf{U}^{I}(\mathbf{k}) = \sum_{J=1}^{N}\mathbf{D}^{xx}_{IJ}(\mathbf{k})\mathbf{U}^{J}(\mathbf{k})
    +\sum_{J=1}^{N}\mathbf{D}^{xp}_{IJ}(\mathbf{k})\mathbf{Q}^{J}(\mathbf{k}),
\end{equation}
where
\begin{eqnarray}
  && \mathbf{D}^{xx}_{IJ}(\mathbf{k}) = \delta_{IJ}\frac{1}{m_I}\frac{\partial^{2}\mathcal{E}}{\partial\mathbf{x}^{I}\partial\mathbf{x}^{I}}\left(\mathcal{B}\right)
  +\sideset{}{'}{\sum}_{j\in \mathcal{L}_{J}}\frac{1}{\sqrt{m_Im_J}}\frac{\partial^{2}\mathcal{E}}{\partial\mathbf{x}^j\partial\mathbf{x}^{I}}\left(\mathcal{B}\right)e^{\mathbbm{i}\mathbf{k}\cdot(\mathbf{X}^j-\mathbf{X}^{I})},
   \nonumber \\
  &&\mathbf{D}^{xp}_{IJ}(\mathbf{k}) = \delta_{IJ}\frac{1}{\sqrt{m_I}}\frac{\partial^{2}\mathcal{E}}{\partial\mathbf{P}^{I}\partial\mathbf{x}^{I}}\left(\mathcal{B}\right)
  +\sideset{}{'}{\sum}_{j\in \mathcal{L}_{J}}\frac{1}{\sqrt{m_I}}\frac{\partial^{2}\mathcal{E}}{\partial\mathbf{P}^j\partial\mathbf{x}^{I}}\left(\mathcal{B}\right)e^{\mathbbm{i}\mathbf{k}\cdot(\mathbf{X}^j-\mathbf{X}^{I})}.
\end{eqnarray}
Similarly, Eq. (\ref{equations-motion}) is simplified to read
\begin{eqnarray}
  && \frac{1}{\sqrt{m_I}}\frac{\partial^{2}\mathcal{E}}{\partial\mathbf{x}^{I}\partial\mathbf{P}^{I}}\left(\mathcal{B}\right)\mathbf{U}^{I}(\mathbf{k})
  +\sum_{J=1}^{N}~\sideset{}{'}{\sum}_{j\in \mathcal{L}_{J}}\frac{1}{\sqrt{m_J}}\frac{\partial^{2}\mathcal{E}}{\partial\mathbf{x}^j\partial\mathbf{P}^{I}}\left(\mathcal{B}\right)e^{\mathbbm{i}\mathbf{k}\cdot(\mathbf{X}^j-\mathbf{X}^{I})}
   \mathbf{U}^{J}(\mathbf{k})  \nonumber \\
  &&~~~+ \frac{\partial^{2}\mathcal{E}}{\partial\mathbf{P}^{I}\partial\mathbf{P}^{I}}\left(\mathcal{B}\right)\mathbf{Q}^{I}(\mathbf{k})
  +\sum_{J=1}^{N}~\sideset{}{'}{\sum}_{j\in \mathcal{L}_{J}}\frac{\partial^{2}\mathcal{E}}{\partial\mathbf{P}^j\partial\mathbf{P}^{I}}\left(\mathcal{B}\right)e^{\mathbbm{i}\mathbf{k}\cdot(\mathbf{X}^j-\mathbf{X}^{I})}
   \mathbf{Q}^{J}(\mathbf{k})=\mathbf{0}.
\end{eqnarray}
Or
\begin{equation}\
    \sum_{J=1}^{N}\mathbf{D}^{px}_{IJ}(\mathbf{k})\mathbf{U}^{J}(\mathbf{k})
    +\sum_{J=1}^{N}\mathbf{D}^{pp}_{IJ}(\mathbf{k})\mathbf{Q}^{J}(\mathbf{k})=\mathbf{0},
\end{equation}
where
\begin{eqnarray}
  && \mathbf{D}^{px}_{IJ}(\mathbf{k}) = \delta_{IJ}\frac{1}{\sqrt{m_I}}\frac{\partial^{2}\mathcal{E}}{\partial\mathbf{x}^{I}\partial\mathbf{P}^{I}}\left(\mathcal{B}\right)
  +\sideset{}{'}{\sum}_{j\in \mathcal{L}_{J}}\frac{1}{\sqrt{m_J}}\frac{\partial^{2}\mathcal{E}}{\partial\mathbf{x}^j\partial\mathbf{P}^{I}}\left(\mathcal{B}\right)e^{\mathbbm{i}\mathbf{k}\cdot(\mathbf{X}^j-\mathbf{X}^{I})},
   \nonumber \\
  &&\mathbf{D}^{pp}_{IJ}(\mathbf{k}) = \delta_{IJ}\frac{\partial^{2}\mathcal{E}}{\partial\mathbf{P}^{I}\partial\mathbf{P}^{I}}\left(\mathcal{B}\right)
  +\sideset{}{'}{\sum}_{j\in \mathcal{L}_{J}} \frac{\partial^{2}\mathcal{E}}{\partial\mathbf{P}^j\partial\mathbf{P}^{I}}\left(\mathcal{B}\right)e^{\mathbbm{i}\mathbf{k}\cdot(\mathbf{X}^j-\mathbf{X}^{I})}.
\end{eqnarray}
Defining
\begin{equation}\
    \mathbf{U}=\left(%
\begin{array}{c}
  \mathbf{U}^1 \\
  \vdots \\
  \mathbf{U}^N \\
\end{array}%
\right),~~~\mathbf{Q}=\left(%
\begin{array}{c}
  \mathbf{Q}^1 \\
  \vdots \\
  \mathbf{Q}^N \\
\end{array}%
\right)
\end{equation}
the linearized equations of motion read
\begin{equation}\
    \mathbf{D}_{xx}(\mathbf{k})\mathbf{U}(\mathbf{k})+\mathbf{D}_{xp}(\mathbf{k})\mathbf{Q}(\mathbf{k})=\omega(\mathbf{k})^2\mathbf{U}(\mathbf{k}),
    ~~~\mathbf{D}_{px}(\mathbf{k})\mathbf{U}(\mathbf{k})+\mathbf{D}_{pp}(\mathbf{k})\mathbf{Q}(\mathbf{k})=\mathbf{0},
\end{equation}
where
\begin{eqnarray}
  && \mathbf{D}_{xx}=\left(%
\begin{array}{ccc}
  \mathbf{D}_{11}^{xx} & \hdots & \mathbf{D}_{1N}^{xx} \\
  \vdots & \ddots & \vdots \\
  \mathbf{D}_{N1}^{xx} & \hdots & \mathbf{D}_{NN}^{xx} \\
\end{array}%
\right),~~~\mathbf{D}_{xp}=\left(%
\begin{array}{ccc}
  \mathbf{D}_{11}^{xp} & \hdots & \mathbf{D}_{1N}^{xp} \\
  \vdots & \ddots & \vdots \\
  \mathbf{D}_{N1}^{xp} & \hdots & \mathbf{D}_{NN}^{xp} \\
\end{array}%
\right), \nonumber \\
  && \nonumber \\
  && \mathbf{D}_{px}=\left(%
\begin{array}{ccc}
  \mathbf{D}_{11}^{px} & \hdots & \mathbf{D}_{1N}^{px} \\
  \vdots & \ddots & \vdots \\
  \mathbf{D}_{N1}^{px} & \hdots & \mathbf{D}_{NN}^{px} \\
\end{array}%
\right),~~~\mathbf{D}_{pp}=\left(%
\begin{array}{ccc}
  \mathbf{D}_{11}^{pp} & \hdots & \mathbf{D}_{1N}^{pp} \\
  \vdots & \ddots & \vdots \\
  \mathbf{D}_{N1}^{pp} & \hdots & \mathbf{D}_{NN}^{pp} \\
\end{array}%
\right).
\end{eqnarray}
Finally, the effective dynamical problem can be written as
\begin{equation}\
    \mathbf{D}(\mathbf{k})\mathbf{U}(\mathbf{k})=\omega(\mathbf{k})^2\mathbf{U}(\mathbf{k}),
\end{equation}
where
\begin{equation}\
    \mathbf{D}(\mathbf{k})=\mathbf{D}_{xx}(\mathbf{k})-\mathbf{D}_{xp}(\mathbf{k})\mathbf{D}_{pp}^{-1}(\mathbf{k})\mathbf{D}_{px}(\mathbf{k}),
\end{equation}
is the effective dynamical matrix. Note that
$\mathbf{D}(\mathbf{k})$ is Hermitian. Denoting the $2N$ eigenvalues
of $\mathbf{D}(\mathbf{k})$ by
$\lambda_i(\mathbf{k})=\omega_i^2(\mathbf{k}),~i=1,...,2N$, free energy of the unit cell is expressed as
\begin{equation}
    \mathcal{F}\left(\{\mathbf{X}^j,\mathbf{\bar{P}}^j\}_{j\in\mathcal{L}},T\right)=\mathcal{E}\left(\{\mathbf{X}^j,\mathbf{\bar{P}}^j\}_{j\in\mathcal{L}}\right)
    +\sum_{\mathbf{k}}\sum_{i=1}^{2N}\left\{\frac{1}{2}\hbar  \omega_i(\mathbf{k})
    +k_BT\ln\left[1-\exp\left(\!-\frac{\hbar \omega_i(\mathbf{k})}{k_BT}\right)\right]\right \}.
\end{equation}
Therefore, for the optimum configuration
$\left\{\mathbf{X}^j,\mathbf{\bar{P}}^j\right\}_{j\in\mathcal{L}}$
at temperature $T$ we should have
\begin{eqnarray}
  && \frac{\partial \mathcal{F}}{\partial \mathbf{X}^j}=\frac{\partial \mathcal{E}}{\partial \mathbf{X}^j}+
  \sum_{\mathbf{k}}\sum_{i=1}^{2N}\left\{\frac{\hbar}{2\omega_i(\mathbf{k})}\left(
    \frac{1}{2}+\frac{1}{\exp\left(\frac{\hbar \omega_i(\mathbf{k})}{k_BT}\right)-1}\right)
    \frac{\partial \omega^2_i(\mathbf{k})}{\partial \mathbf{X}^j}\right\}=\mathbf{0}, \\
  && \frac{\partial \mathcal{F}}{\partial \mathbf{\bar{P}}^j}=\frac{\partial \mathcal{E}}{\partial \mathbf{\bar{P}}^j}+
  \sum_{\mathbf{k}}\sum_{i=1}^{2N}\left\{\frac{\hbar}{2\omega_i(\mathbf{k})}\left(
    \frac{1}{2}+\frac{1}{\exp\left(\frac{\hbar \omega_i(\mathbf{k})}{k_BT}\right)-1}\right)
    \frac{\partial \omega^2_i(\mathbf{k})}{\partial \mathbf{\bar{P}}^j}\right\}=\mathbf{0},
\end{eqnarray}
where the derivatives of eigenvalues are given by
\begin{equation}
  \frac{\partial \omega^2_{i} \left(\mathbf{k}\right)}{\partial \mathbf{X}^j}=
  \sum_{\alpha , \beta=1}^{2N}V^{*}_{\alpha i}\left(\mathbf{k}\right)
  \frac{\partial D_{\alpha \beta}  \left(\mathbf{k}\right)} {\partial \mathbf{X}^j}V_{\beta i}\left(\mathbf{k}\right),~~~
  \frac{\partial \omega^2_{i} \left(\mathbf{k}\right)}{\partial \mathbf{\bar{P}}^j}=
  \sum_{\alpha , \beta=1}^{2N}V^{*}_{\alpha i}\left(\mathbf{k}\right)
  \frac{\partial D_{\alpha \beta}  \left(\mathbf{k}\right)} {\partial \mathbf{\bar{P}}^j}V_{\beta i}\left(\mathbf{k}\right),
\end{equation}
where
$\mathbf{V}\left(\mathbf{k}\right)=\left[V_{\alpha\beta}\left(\mathbf{k}\right)\right]\in
\mathbb{R}^{2N \times 2N}$ is the matrix of the eigenvectors of
$\mathbf{D}(\mathbf{k})=\left[D_{\alpha\beta}\left(\mathbf{k}\right)\right]$,
with $D_{\alpha\beta}$ normalized to unity.

\paragraph{Dynamical Matrix for the Defective Lattice} In the case of
a defective lattice we consider interactions of order $m$, i.e., we write
\begin{equation}
    \mathcal{L}_i=\bigsqcup_{\alpha=-m}^{m}\bigsqcup_{I=1}^{N}\mathcal{L}_{I\alpha},
\end{equation}
where $\mathcal{L}_i$ is the neighboring set of the atom $i$. The
equations of motion (\ref{equations-motion3}) and
(\ref{equations-motion4}) become
\begin{eqnarray}
  \omega(\mathbf{k})^2\mathbf{U}^{I\alpha}(\mathbf{k}) &=& \sum_{J=1}^{N}\sum_{\beta=-m}^{m}\mathbf{D}^{xx}_{I\alpha J\beta}(\mathbf{k})\mathbf{U}^{J\beta}(\mathbf{k})
    +\sum_{J=1}^{N}\sum_{\beta=-m}^{m}\mathbf{D}^{xp}_{I\alpha J\beta}(\mathbf{k})\mathbf{Q}^{J\beta}(\mathbf{k}), \\
  \mathbf{0} &=& \sum_{J=1}^{N}\sum_{\beta=-m}^{m}\mathbf{D}^{px}_{I\alpha J\beta}(\mathbf{k})\mathbf{U}^{J\beta}(\mathbf{k})
    +\sum_{J=1}^{N}\sum_{\beta=-m}^{m}\mathbf{D}^{pp}_{I\alpha
    J\beta}(\mathbf{k})\mathbf{Q}^{J\beta}(\mathbf{k}).
\end{eqnarray}
Defining
\begin{equation}\
    \mathbf{U}_{\alpha}=\left(%
\begin{array}{c}
  \mathbf{U}^{1\alpha} \\
  \vdots \\
  \mathbf{U}^{N\alpha} \\
\end{array}%
\right)\in\mathbb{R}^{2N},~~~\mathbf{Q}_{\alpha}=\left(%
\begin{array}{c}
  \mathbf{Q}^{1\alpha} \\
  \vdots \\
  \mathbf{Q}^{N\alpha} \\
\end{array}%
\right)\in\mathbb{R}^{2N},
\end{equation}
we can write the equations of motion as follows
\begin{eqnarray}
  \omega(\mathbf{k})^2\mathbf{U}_{\alpha}(\mathbf{k}) &=& \sum_{\beta=-m}^{m}\mathbf{A}^{xx}_{\alpha \left(\alpha+\beta\right)}(\mathbf{k})\mathbf{U}_{\left(\alpha+\beta\right)}(\mathbf{k})
    +\sum_{\beta=-m}^{m}\mathbf{A}^{xp}_{\alpha \left(\alpha+\beta\right)}(\mathbf{k})\mathbf{Q}_{\left(\alpha+\beta\right)}(\mathbf{k}), \\
  \mathbf{0} &=& \sum_{\beta=-m}^{m}\mathbf{A}^{px}_{\alpha \left(\alpha+\beta\right)}(\mathbf{k})\mathbf{U}_{\left(\alpha+\beta\right)}(\mathbf{k})
    +\sum_{\beta=-m}^{m}\mathbf{A}^{pp}_{\alpha
    \left(\alpha+\beta\right)}(\mathbf{k})\mathbf{Q}_{\left(\alpha+\beta\right)}(\mathbf{k}),
\end{eqnarray}
where
\begin{eqnarray}
  && \mathbf{A}^{\ast\star}_{\alpha \beta}=\left(%
\begin{array}{ccc}
  \mathbf{D}_{1\alpha1\beta}^{\ast\star} & \hdots & \mathbf{D}_{1\alpha N\beta}^{\ast\star} \\
  \vdots & \ddots & \vdots \\
  \mathbf{D}_{N\alpha1\beta}^{\ast\star} & \hdots & \mathbf{D}_{N\alpha N\beta}^{\ast\star} \\
\end{array}%
\right)\in\mathbb{R}^{2N\times 2N}~~~\ast,\star=x,p.
\end{eqnarray}
Let us consider only a finite number of equivalence classes around
the defect, i.e., we assume that $-C\leq\alpha\leq C$. Therefore,
the approximating finite system has the following governing
equations
\begin{eqnarray}
  && \mathbf{D}_{xx}(\mathbf{k})\mathbf{U}(\mathbf{k})+\mathbf{D}_{xp}(\mathbf{k})\mathbf{Q}(\mathbf{k}) = \omega(\mathbf{k})^2\mathbf{U}(\mathbf{k}), \\
  && \mathbf{D}_{px}(\mathbf{k})\mathbf{U}(\mathbf{k})+\mathbf{D}_{pp}(\mathbf{k})\mathbf{Q}(\mathbf{k}) = \mathbf{0},
\end{eqnarray}
where
\begin{equation}\
    \mathbf{U}(\mathbf{k})=\left(%
\begin{array}{c}
  \mathbf{U}_{-C} \\
  \vdots \\
  \mathbf{U}_{C} \\
\end{array}%
\right)\in\mathbb{R}^{M},~~~\mathbf{Q}(\mathbf{k})=\left(%
\begin{array}{c}
  \mathbf{Q}_{-C} \\
  \vdots \\
  \mathbf{Q}_{C} \\
\end{array}%
\right) \in\mathbb{R}^{M},
\end{equation}
\begin{eqnarray}
  && \mathbf{D}_{\ast\star}(\mathbf{k})=\left(%
\begin{array}{ccc}
  \mathbb{D}^{\ast\star}_{\left(-C\right)\left(-C\right)} & \hdots & \mathbb{D}^{\ast\star}_{\left(-C\right)C} \\
  \vdots & \ddots & \vdots \\
  \mathbb{D}^{\ast\star}_{C\left(-C\right)} & \hdots & \mathbb{D}^{\ast\star}_{CC} \\
\end{array}%
\right)\in\mathbb{R}^{M\times M},~~~
\mathbb{D}^{\ast\star}_{\alpha\beta}= \left\{ %
\begin{array}{c}
  \mathbf{A}^{\ast\star}_{\alpha\beta}~~~~~~~~~|\alpha-\beta|\leq m, \\
  \\
  \mathbf{0}_{2N\times2N} ~~~~|\alpha-\beta|> m.\\
\end{array}%
\right.,
\end{eqnarray}
where $M=2N\times(2C+1)$ and $\ast,\star=x,p$.
Now the effective dynamical problem can be written as
\begin{equation}\
    \mathbf{D}(\mathbf{k})\mathbf{U}(\mathbf{k})=\omega(\mathbf{k})^2\mathbf{U}(\mathbf{k}),
\end{equation}
where
\begin{equation}\ \label{dynamicalmatrix}
    \mathbf{D}(\mathbf{k})=\mathbf{D}_{xx}(\mathbf{k})-\mathbf{D}_{xp}(\mathbf{k})\mathbf{D}_{pp}^{-1}(\mathbf{k})\mathbf{D}_{px}(\mathbf{k}),
\end{equation}
is the effective dynamical matrix. Note that
$\mathbf{D}(\mathbf{k})$ is Hermitian and has $M$ real eigenvalues.
The free energy of the unit cell is expressed as
\begin{equation}
    \mathcal{F}\left(\{\mathbf{X}^j,\mathbf{\bar{P}}^j\}_{j\in\mathcal{L}},T\right)=\mathcal{E}\left(\{\mathbf{X}^j,\mathbf{\bar{P}}^j\}_{j\in\mathcal{L}}\right)
    +\sum_{\mathbf{k}}\sum_{i=1}^{M}\left\{\frac{1}{2}\hbar  \omega_i(\mathbf{k})
    +k_BT\ln\left[1-\exp\left(\!-\frac{\hbar \omega_i(\mathbf{k})}{k_BT}\right)\right]\right\}.
\end{equation}
For the optimum structure
$\left\{\mathbf{X}^j,\mathbf{\bar{P}}^j\right\}_{j\in\mathcal{L}}$
at temperature $T$ we have
\begin{eqnarray}
  && \frac{\partial \mathcal{F}}{\partial \mathbf{X}^j}=\frac{\partial \mathcal{E}}{\partial \mathbf{X}^j}+
  \sum_{\mathbf{k}}\sum_{i=1}^{M}\left\{\frac{\hbar}{2\omega_i(\mathbf{k})}\left(
    \frac{1}{2}+\frac{1}{\exp\left(\frac{\hbar \omega_i(\mathbf{k})}{k_BT}\right)-1}\right)
    \frac{\partial \omega^2_i(\mathbf{k})}{\partial \mathbf{X}^j}\right\}=\mathbf{0}, \\
  && \frac{\partial \mathcal{F}}{\partial \mathbf{\bar{P}}^j}=\frac{\partial \mathcal{E}}{\partial \mathbf{\bar{P}}^j}+
  \sum_{\mathbf{k}}\sum_{i=1}^{M}\left\{\frac{\hbar}{2\omega_i(\mathbf{k})}\left(
    \frac{1}{2}+\frac{1}{\exp\left(\frac{\hbar \omega_i(\mathbf{k})}{k_BT}\right)-1}\right)
    \frac{\partial \omega^2_i(\mathbf{k})}{\partial \mathbf{\bar{P}}^j}\right\}=\mathbf{0},
\end{eqnarray}
where the derivatives of eigenvalues are given by
\begin{equation}
  \frac{\partial \omega^2_{i} \left(\mathbf{k}\right)}{\partial \mathbf{X}^j}=
  \sum_{\alpha , \beta=1}^{M}V^{*}_{\alpha i}\left(\mathbf{k}\right)
  \frac{\partial D_{\alpha \beta}  \left(\mathbf{k}\right)} {\partial \mathbf{X}^j}V_{\beta i}\left(\mathbf{k}\right),
  ~~~
  \frac{\partial \omega^2_{i} \left(\mathbf{k}\right)}{\partial \mathbf{\bar{P}}^j}=
  \sum_{\alpha , \beta=1}^{M}V^{*}_{\alpha i}\left(\mathbf{k}\right)
  \frac{\partial D_{\alpha \beta}  \left(\mathbf{k}\right)} {\partial \mathbf{\bar{P}}^j}V_{\beta i}\left(\mathbf{k}\right),
\end{equation}
where
$\mathbf{V}\left(\mathbf{k}\right)=\left[V_{\alpha\beta}\left(\mathbf{k}\right)\right]\in
\mathbb{R}^{M \times M}$ is the matrix of the eigenvectors of
$\mathbf{D}(\mathbf{k})=\left[D_{\alpha\beta}\left(\mathbf{k}\right)\right]$,
with $D_{\alpha\beta}$ normalized to unity.

\section{Temperature-Dependent Structure of $180^{\circ}$ Domain Walls in a 2-D Lattice of Dipoles}

\begin{figure}[hbt]
\begin{center}
\includegraphics[scale=0.7,angle=0]{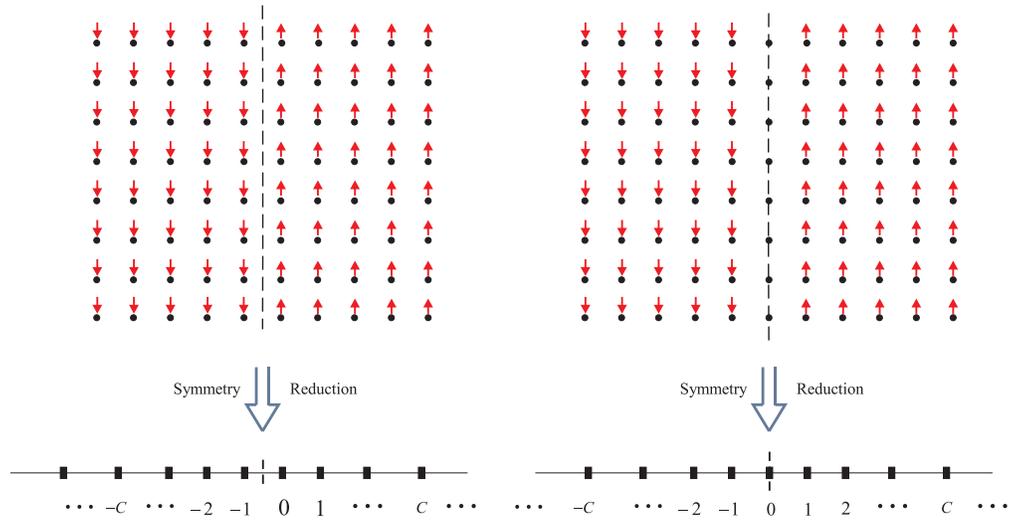}
\end{center}
\vspace*{-0.2in}\caption{\footnotesize Reference configurations for
the $180^{\circ}$ domain walls in the 2-D lattice of dipoles, their
symmetry reduction and their reduced lattices. Left Panel: Type I,
Right Panel: Type II.} \label{180}
\end{figure}

To demonstrate the capabilities of our lattice dynamics technique,
here we consider a simple example of $180^{\circ}$ domain walls
shown in Fig. \ref{180}. In these $180^{\circ}$ domain walls,
polarization vector changes from $-\mathbf{P}_0$ on the left side of
the domain wall to $\mathbf{P}_0$ on the right side of the domain
wall. We consider two types of domain walls: Type I and Type II. In
Type I (the left configuration) the domain wall is not a
crystallographic line,  but it passes through some atoms in Type II
(the right configuration). We are interested in the structure of the
defective lattice close to the domain wall at a finite temperature
$T$. In these examples, each equivalent class is a set of atoms
lying on a line parallel to the domain wall, i.e., we have a
defective crystal with a 1-D symmetry reduction. The static
configurations for Type I domain wall, $\mathcal{B}_0$, was computed
in \citep{Yavari2005a}. Here we consider the static equilibrium
configurations as the initial reference configurations. For index
$n\in\mathbb{Z}$ in the reduced lattice (see Fig. \ref{180}), the
vectors of unknowns are $\mathbf{U}_n,\mathbf{Q}_n\in\mathbb{R}^2$.
Because of symmetry, we only consider the right half of the lattices
and because the effective potential is highly localized
\citep{Yavari2005a}, for calculation of the stiffness matrices, we
assume that a given unit cell interacts only with its nearest
neighbor equivalence classes, i.e., we consider interactions of
order $m=1$.  Note that this choice of $m$ only affects the harmonic
solutions; the final anharmonic solutions are not affected by this
choice. For our numerical calculations we choose $N=280$ atoms in
each equivalence class as the results are independent of $N$ for
larger $N$. Note that for force calculations we consider all the
atoms within a specific cut-off radius $R_c$. Here, we use $R_c
=140a$, where $a$ is the lattice parameter in the nominal
configuration.

For minimizing the free energy, first one should calculate the
effective dynamical matrix according to Eq. (\ref{dynamicalmatrix}).
The calculations of this matrix for the two configurations are
similar. For example, in configuration I due to symmetry we have
$\mathbf{U}_{-1}=-\mathbf{U}_{0}$. Also we consider the temperature-dependent bulk
configuration as the far-field condition, i.e., we assume
$\mathbf{U}_{\alpha}=\mathbf{U}_{C}$ for $\alpha \geq C+1$. Our
numerical experiments show that choosing $C=35$ would be enough to
capture the structure of the atomic displacements near the defect,
so we use $C=35$ in what follows. For the right half of the
defective lattice we have
\begin{eqnarray}
  && \mathbf{D}_{\ast\star}=\left(%
\begin{array}{ccccccc}
  \mathbf{E}^{\ast\star}_{0} & \mathbf{D}^{\ast\star}_{01} & \mathbf{0}_{2\times2} & \hdots & \mathbf{0}_{2\times2} & \mathbf{0}_{2\times2} & \mathbf{0}_{2\times2} \\
  \mathbf{D}^{\ast\star}_{10} & \mathbf{D}^{\ast\star}_{11} & \mathbf{D}^{\ast\star}_{12} & \hdots & \mathbf{0}_{2\times2} & \mathbf{0}_{2\times2} & \mathbf{0}_{2\times2} \\
  \mathbf{0}_{2\times2} & \mathbf{D}^{\ast\star}_{21} & \mathbf{D}^{\ast\star}_{22} & \hdots & \mathbf{0}_{2\times2} & \mathbf{0}_{2\times2} & \mathbf{0}_{2\times2} \\
  \vdots & \vdots & \vdots & \ddots & \vdots & \vdots & \vdots \\
  \mathbf{0}_{2\times2} & \mathbf{0}_{2\times2} & \mathbf{0}_{2\times2} & \hdots & \mathbf{D}^{\ast\star}_{\left(C-2\right)\left(C-2\right)}
   & \mathbf{D}^{\ast\star}_{\left(C-2\right)\left(C-1\right)} & \mathbf{0}_{2\times2} \\
  \mathbf{0}_{2\times2} & \mathbf{0}_{2\times2} & \mathbf{0}_{2\times2} & \hdots & \mathbf{D}^{\ast\star}_{\left(C-1\right)\left(C-2\right)}
   & \mathbf{D}^{\ast\star}_{\left(C-1\right)\left(C-1\right)} & \mathbf{D}^{\ast\star}_{\left(C-1\right)C} \\
  \mathbf{0}_{2\times2} & \mathbf{0}_{2\times2} & \mathbf{0}_{2\times2} & \hdots & \mathbf{0}_{2\times2}
   & \mathbf{D}^{\ast\star}_{C\left(C-1\right)} & \mathbf{F}^{\ast\star}_{C} \\
\end{array}%
\right)\in\mathbb{R}^{S\times S},
\end{eqnarray}
where $S=2\left(C+1\right)$,
\begin{equation}
    \mathbf{E}^{\ast\star}_0=\mathbf{D}^{\ast\star}_{00}-\mathbf{D}^{\ast\star}_{0\left(-1\right)}~~~~~
    \text{and}~~~~~
    \mathbf{F}^{\ast\star}_C=\mathbf{D}^{\ast\star}_{CC}+\mathbf{D}^{\ast\star}_{C\left(C+1\right)}~~~~~~~\ast,\star=x,p.
\end{equation}
Now one can use the above matrices to calculate the effective
dynamical matrix. Note that as a consequence of considering
interaction of order $m$, the dynamical matrix will be sparse, i.e.,
only a small number of elements are nonzero. As the dimension of the
system increases, sparsity can be very helpful in the numerical
computations \citep{PressTVF1989}.
\begin{figure}[hbt]
\begin{center}
\includegraphics[scale=0.35,angle=0]{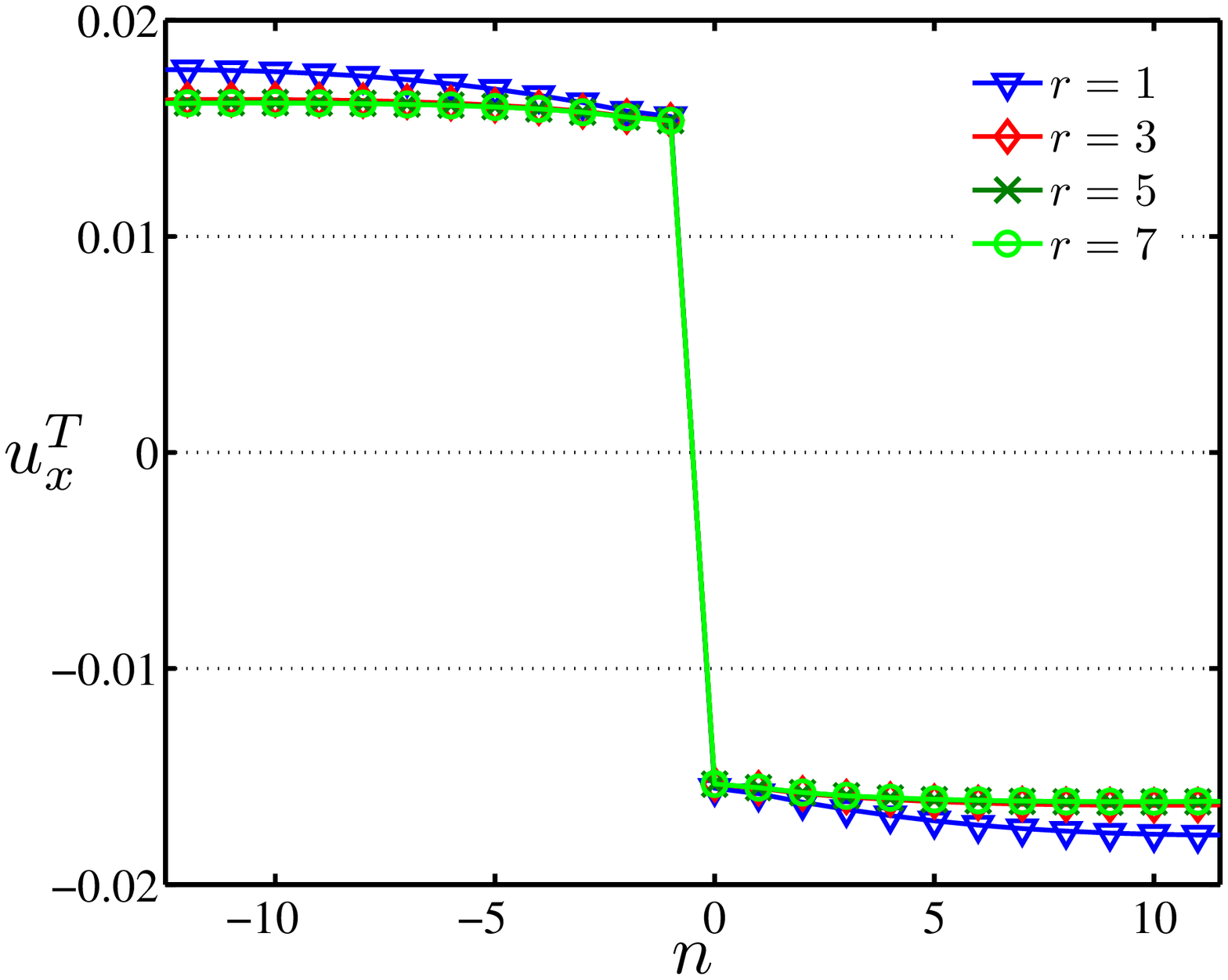}
\includegraphics[scale=0.35,angle=0]{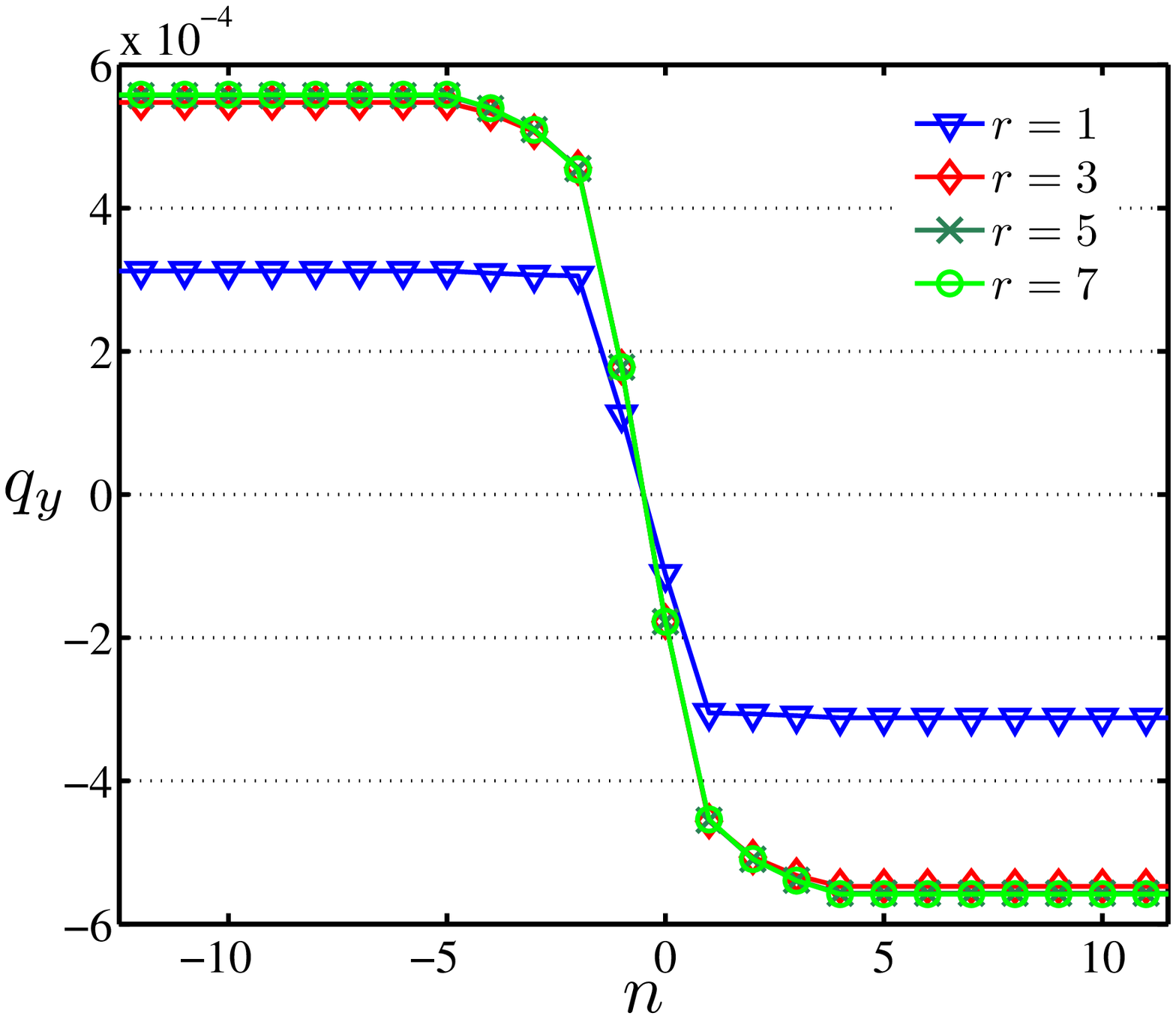}
\end{center}
\vspace*{-0.2in}\caption{\footnotesize Position and polarization displacements for
Type I domain wall ($\bar{T} =5$) obtained by choosing different number of
$\mathbf{k}$-points ($r$) in the integration over the first
Brillouin zone.} \label{T5_r}
\end{figure}

As was mentioned earlier, we will consider only the static part of
the free energy to build the Hessian for the initial iteration and
then update the Hessian using the BFGS algorithm in each step. To
calculate the gradient of the free energy we need the third
derivatives of the potential energy. These can be calculated using
following relation
\begin{equation}
    \frac{\partial \mathbf{D}}{\partial \boldsymbol{\Xi}}=\frac{\partial \mathbf{D}_{xx}}{\partial \boldsymbol{\Xi}}-
    \frac{\partial \mathbf{D}_{xp}}{\partial \boldsymbol{\Xi}}\mathbf{D}^{-1}_{pp}\mathbf{D}_{px}+
    \mathbf{D}_{xp}\mathbf{D}^{-1}_{pp}\frac{\partial \mathbf{D}_{pp}}{\partial \boldsymbol{\Xi}}\mathbf{D}^{-1}_{pp}\mathbf{D}_{px}-
    \mathbf{D}_{xp}\mathbf{D}^{-1}_{pp}\frac{\partial \mathbf{D}_{px}}{\partial \boldsymbol{\Xi}}~~~~~~~~~~\boldsymbol{\Xi}=\mathbf{X}^i,\mathbf{\bar{P}}^i.
\end{equation}
To obtain these third derivatives one can use the translation
invariance relations (\ref{invar1}) and (\ref{invar2}) to simplify
the calculations. For example, we can write
\begin{equation}
     \frac{\partial^{3}\mathcal{E}}{\partial\mathbf{x}^{i}\partial\mathbf{x}^{i}\partial\mathbf{x}^{i}}\left(\mathcal{B}\right)=
    -\sideset{}{'}{\sum}_{j\in\mathcal{L}_{i}}\frac{\partial^{3}\mathcal{E}}{\partial\mathbf{x}^j\partial\mathbf{x}^{i}\partial\mathbf{x}^{i}}\left(\mathcal{B}\right),
\end{equation}
where a prime means that we exclude $j=i$ from the summation.

The dimensionalized temperature $\bar{T}$ and dimensionalized mass $\bar{m}$ correspond to the choice $\hbar = k_B = 10^{-3}$.\footnote{ We select these values to be able to work with temperatures that are comparable with real temperature values.} and work with normalized . To obtain the static equilibrium
configuration and also in dynamic calculations we use $a=1.0$, $P_0
=1.0$, $\epsilon=0.125$ , $K_A =2.0$ and $\bar{m} = 10^4$. In what
follows convergence tolerance for $\sqrt{\boldsymbol{\nabla}
\mathcal{F}\cdot \boldsymbol{\nabla} \mathcal{F}^{\textsf{T}}}$ is
$10^{-5}$. Using this value for convergence tolerance, solutions
converge after ten to twenty iterations. In Fig. \ref{T5_r} we plot $u^T_x$ and $q_y$ for Type I domain wall
and $\bar{T} =5$ for different number of $\mathbf{k}$-points ($r$)
in the first Brillouin zone. Here $u^T_x$ is the diplacement of the
lattice with respect to the nominal configuration at temperature
$\bar{T}$.\footnote{Note that as temperature increases, lattice
parameters change. A temperature-dependent nominal configuration is
what is shown in Fig. \ref{180} but with the bulk lattice parameters
at that temperature.} For numerical integrations over the first Brillouin zone we use the special points
introduced in \citep{MonkhorstPack1976}. For the case $r=1$ we set
$\mathbf{k}=0$, i.e., we assume that all of the atoms in a
particular equivalence class vibrate with the same phase. As can be
seen in these figures, displacements converge quickly by selecting
$r=7$ $\mathbf{k}$-points in the first Brillouin zone, so in what
follows we set $r=7$.

\begin{figure}[hbt]
\begin{center}
\includegraphics[scale=0.35,angle=0]{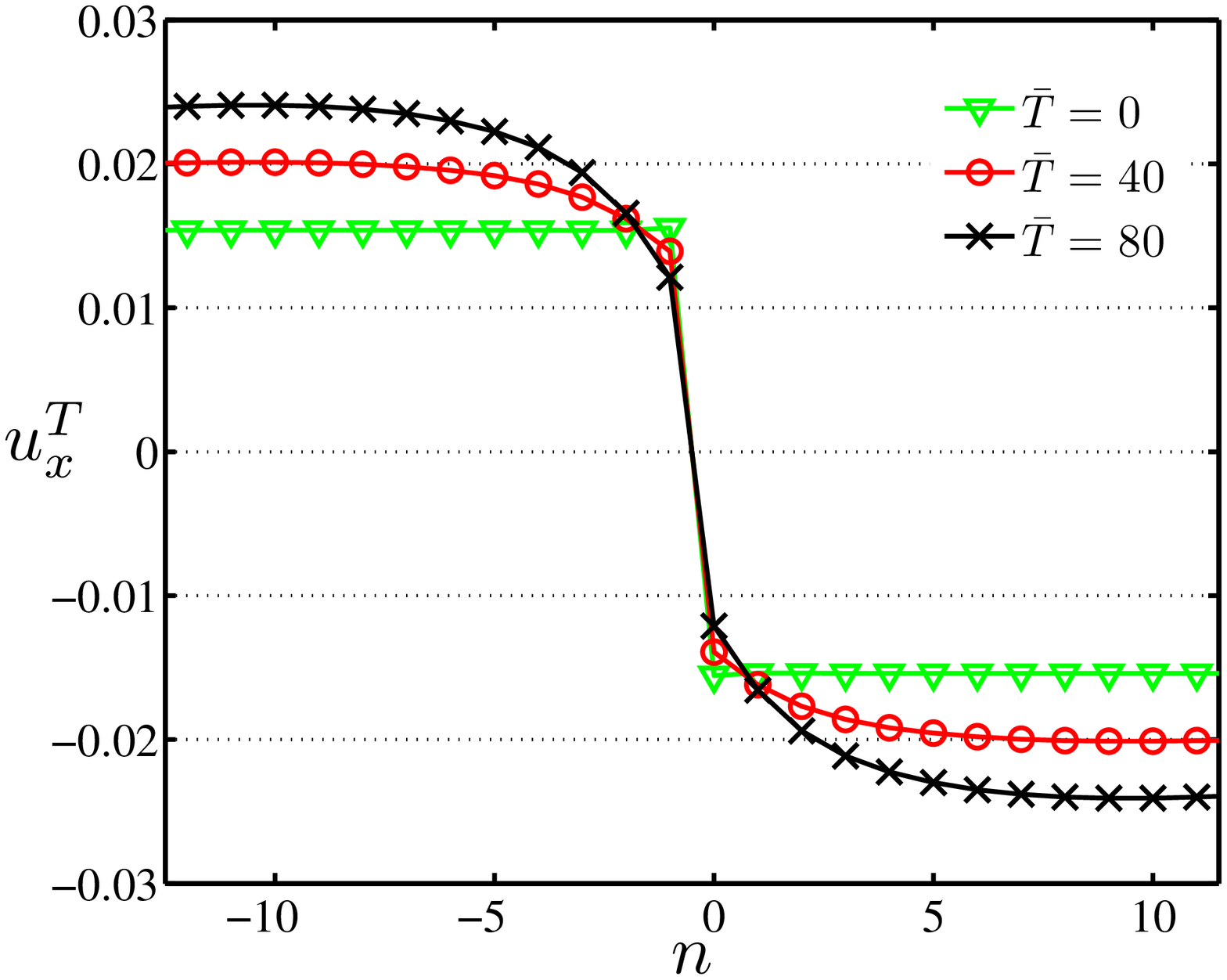}
\includegraphics[scale=0.35,angle=0]{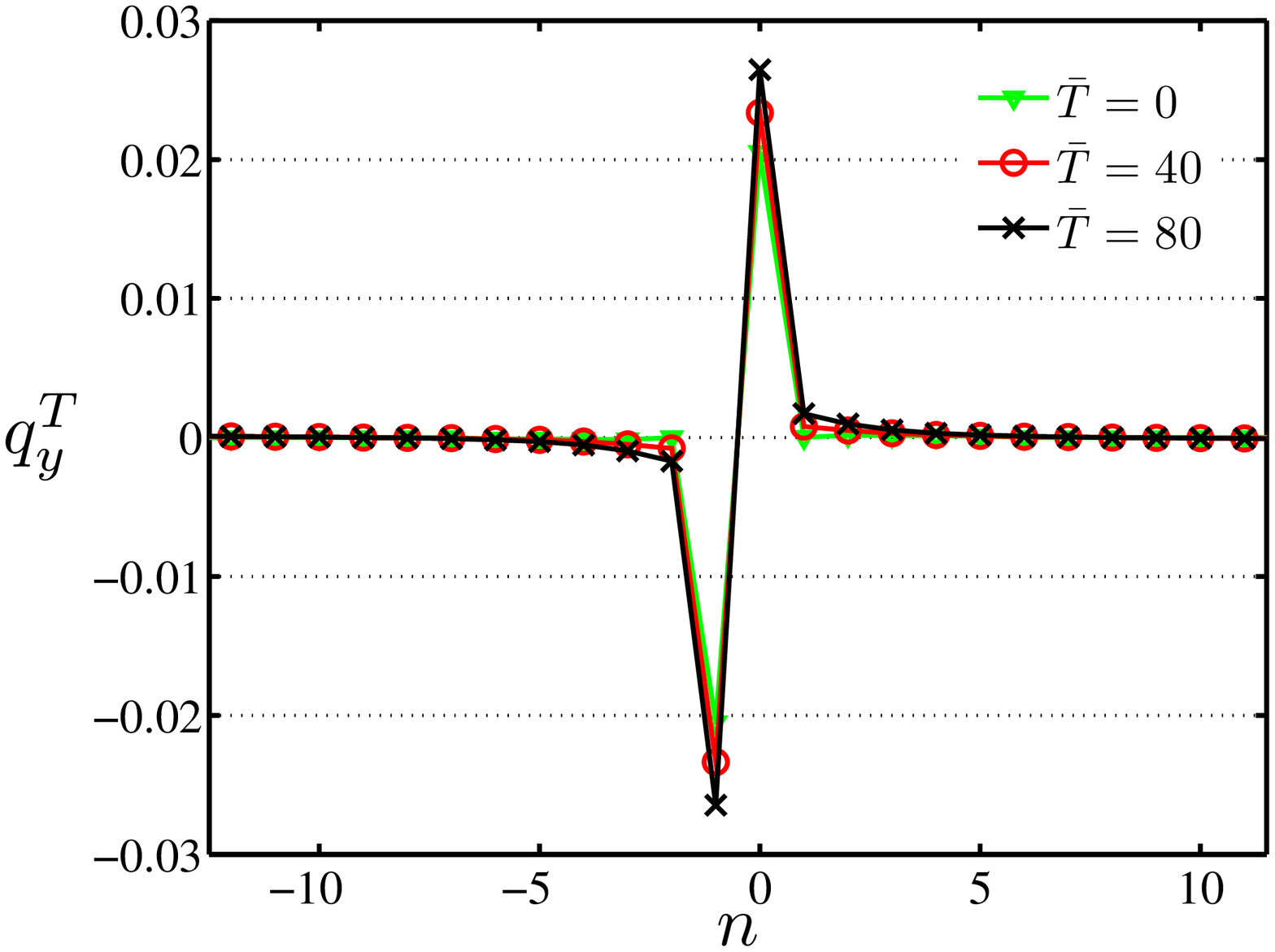}
\end{center}
\vspace*{-0.2in}\caption{\footnotesize Position and polarization displacements of Type I domain wall with respect to the temperature-dependent nominal configurations. }\label{Tdis}
\end{figure}

\begin{figure}[hbt]
\begin{center}
\includegraphics[scale=0.35,angle=0]{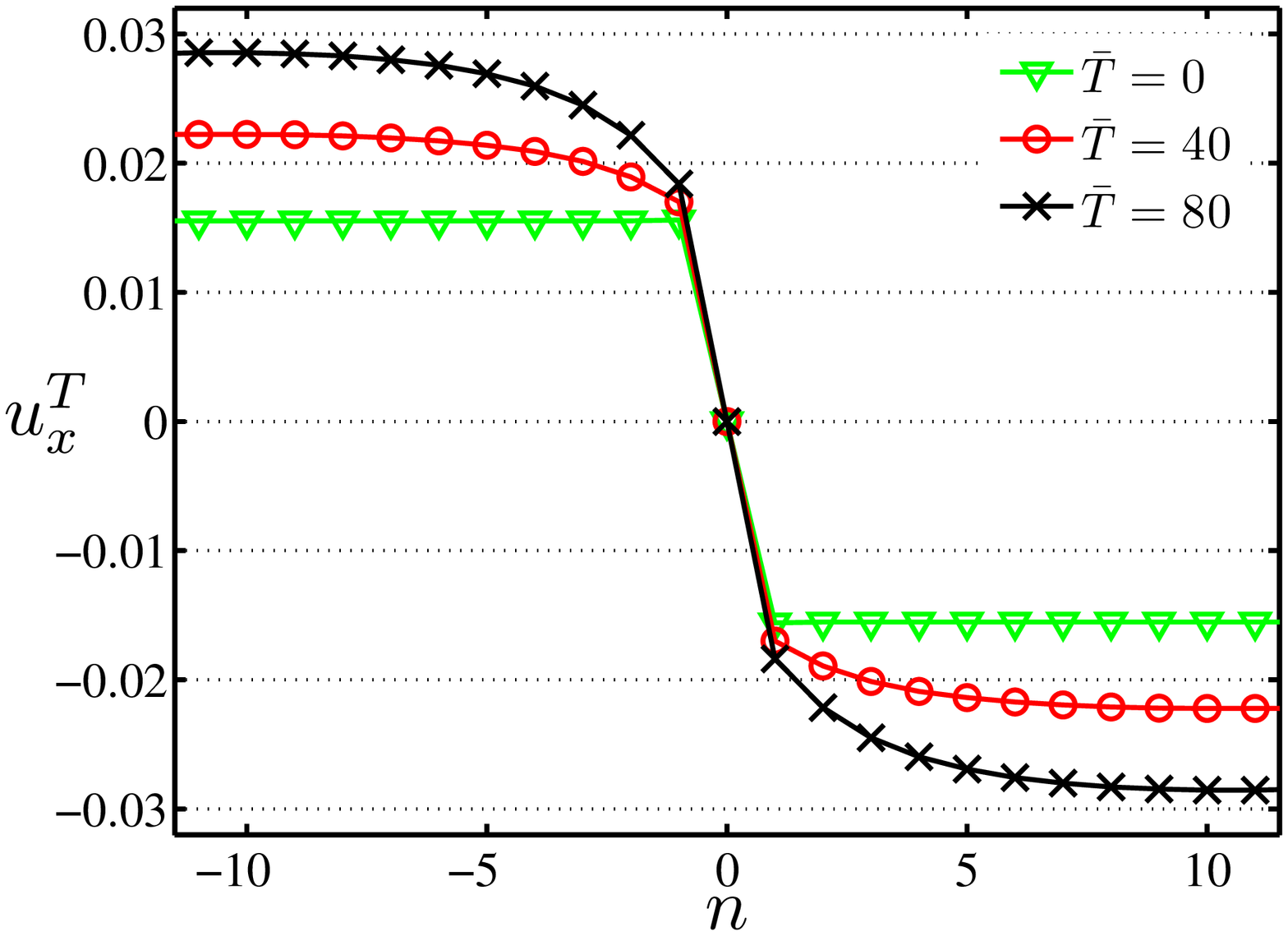}
\includegraphics[scale=0.35,angle=0]{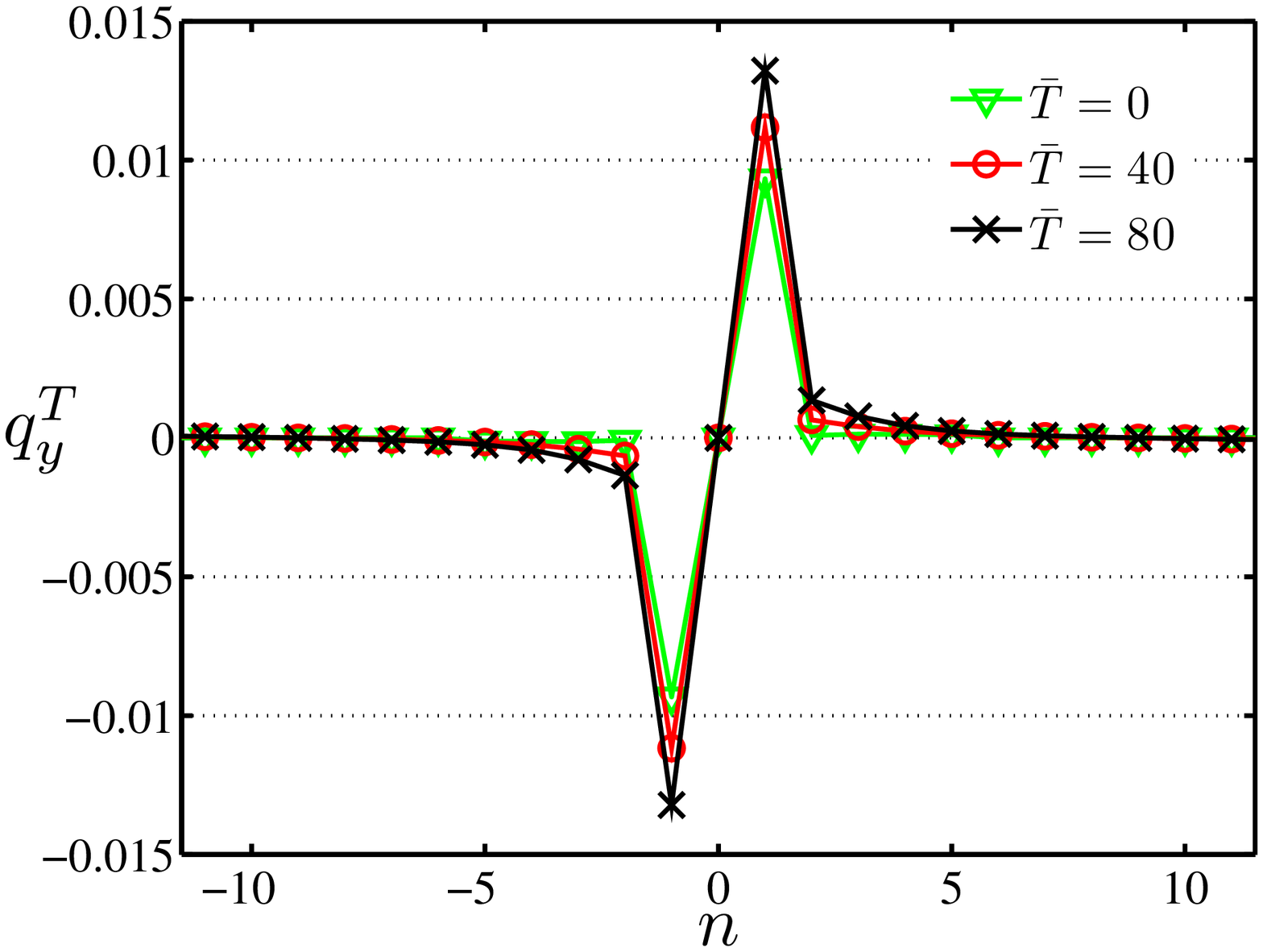}
\end{center}
\vspace*{-0.2in}\caption{\footnotesize Position and polarization displacements of Type II domain wall with respect to the temperature-dependent nominal configurations. }\label{Tdis2}
\end{figure}

Figs. \ref{Tdis} and \ref{Tdis2} show the variations of
displacements with temperature for the two domain walls. As
temperature increases we cannot use the static equilibrium
configuration as the reference configuration for calculating
$\mathbf{H}_0$. Instead, we use the equilibrium configuration of a
smaller temperature to obtain $\mathbf{H}_0$. Here, we use steps
equal to $\Delta \bar{T}=5$. In other words, for calculating the
structure of a domain wall at $\bar{T}=30$, for example, we use the
structure at $\bar{T}=25$ as the initial configuration. We see that
the lattice statics solution and the lattice configuration at $T=0$
obtained by the free energy minimization have a small difference.
Such differences are due to the zero-point motions; the lattice
statics method ignores the quantum effects. It is a well known fact
that zero-point motions can have significant effects in some systems
\citep{KohanoffAndreoniParrinello1992}. Note that polarization near
the domain wall increases with temperature. Also as it is expected,
the lattice expands by increasing the temperature.

\begin{figure}[hbt]
\begin{center}
\includegraphics[scale=0.35,angle=0]{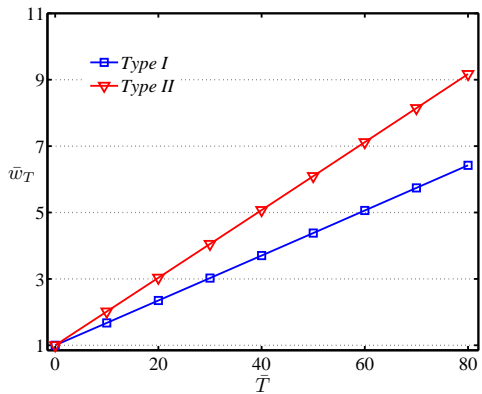}
\end{center}
\vspace*{-0.2in}\caption{\footnotesize Variation of the
$180^{\circ}$ domain wall thickness with temperature.} \label{w_t}
\end{figure}

Only a few layers around the domain
wall are distorted; the rest of the lattice is displaced rigidly. As
we see in Fig. \ref{w_t}, the domain wall thickness for both
configurations increases as temperature increases. In this figure
$\bar{w}_T = w_T / w_0$, where $w_0$ is the domain wall thickness at
$\bar{T}=0$. Note also that in this temperature range $\bar{w}_T$ increases linearly with $\bar{T}$. This qualitatively agrees with experimental observations for PbTiO$_3$
in the low temperature regime \citep{FoethStadelmannRobert2007}. \citet{FoethStadelmannRobert2007} observed that domain wall
thickness increases with temperature. What they measured was an
average domain wall thickness. Note that domain wall thickness
cannot be defined uniquely very much like boundary layer thickness
in fluid mechanics. Here, domain wall thickness is by definition the
region that is affected by the domain wall, i.e. those layers that
are distorted. One can use definitions like the $99\%$-thickness in
fluid mechanics and define the domain wall thickness as the length
of the region that has $99\%$ of the far field rigid translation
displacement. What is important is that no matter what definition is
chosen, domain wall ``thickness" increases by increasing temperature.

Our calculations show that by increasing the mass of the atoms both position and polarization
displacements decrease. However, variations of displacements with
respect to mass is very small. For example, by increasing mass from
$\bar{m} = 10^4$ to $\bar{m} = 10^6$ at $\bar{T} =10$, displacements
decrease by less than $0.1\%$.

\section{Concluding Remarks}

In this paper we extended the classical method of lattice dynamics
to defective crystals. The motivation for developing such a
technique is to semi-analytically obtain the finite-temperature
structure of defects in crystalline solids at low temperatures.
Our technique exploits partial symmetries of defects. We worked
out examples of defects in a 2-D lattice of interacting dipoles. We
obtained the finite-temperature structure of two $180^{\circ}$ domain
walls. We observed that using our simple model potential, increasing
temperature domain walls thicken. This is in agreement
with experimental results for ferroelectric domain walls in
PbTiO$_3$. This technique can be used for many physically
important material systems. Extending the present calculations for
$180^{\circ}$ domain walls in PbTiO$_3$ will be the subject of a
future work.

\appendix
\section{The Ensemble Theories}

There are different ensemble theories for calculating the
thermodynamical properties of systems from the statistical mechanics
point of view. In this appendix, we consider micro canonical and
canonical ensemble theories and discuss the relation between them. In particular, we will see that the free energy minimization
discussed in this paper is equivalent to finding the most probable energy
at the given temperature. For more detailed discussions see
\cite{Pathria1996}.

\subsection{Micro Canonical Ensemble Theory}

From thermodynamical
considerations, it is known that by specifying the limited number of
properties of a system, one can determine all the other properties. In principle, any physical system, i.e., any macro system,
consists of many smaller subsystems. Therefore, we can consider properties of
each macro system as \textit{macrostates} specified by the
properties of these subsystems that are called
\textit{microstates}. Note that by a microstate we mean a set of
values associated to each subsystem of a system. For example,
consider an isolated system with energy $E$ and volume $V$ that
consists of $N$ non-interacting particles with energies $\epsilon_{i}$,
$i=1,2,\ldots,N$. Now each n-topple
$(\epsilon_{1},\ldots,\epsilon_{i})$ satisfying
\begin{eqnarray}
\label{eq:energy}  \sum_{i=1}^{N}\epsilon_i = E,
\end{eqnarray}
would represent a microstate of this system.

Obviously, there may exist several microstates that are associated
to the same macrostate. Let $\Omega(E,N,V)$ denote the number of
microstates associated with the given macrostate $(E,N,V)$. We assume that for an isolated system, (i) all microstate compatible with the given macrostates are equally probable, and (ii) equilibrium corresponds to the
macrostate having the largest number of microstates. Let $S$ and $k_B$
denote the entropy of a system and Boltzmann constant, respectively.
Then one can show that the above two assumptions and setting
\begin{eqnarray}
\label{eq:entropy}  S = k_B\ln(\Omega),
\end{eqnarray}
yields the equality of temperatures for systems that are in
thermodynamical equilibrium. Note that (\ref{eq:entropy}) provides
the fundamental relation between thermodynamics and statistical
mechanics. Once $S$ is obtained, the derivation of other
thermodynamical quantities would be a straight forward task.

\subsection{Canonical Ensemble Theory}

In practice, we never have an
isolated system and even if we have such a system, it is hard to
measure the total energy of the system. This means that it is more convenient to develop a statistical mechanics formalism that does not use $E$ as an independent
variable. It is relatively easy to control the temperature
of a system, i.e. we can always put the system in contact with a
heat bath at temperature $T$. Thus, it is natural to choose
$T$ instead of $E$.

Let a system be in equilibrium with a heat bath at temperature $T$
\footnote{We assume systems can only exchange energy.}. In principle,
the energy of the system at any instant of time can be equal to
any energy level of the system. As a matter of fact, one can show that the
probability of a system being in the energy level $P_r$ is
equal to
\begin{eqnarray}
\label{eq:prob}  P_r =
\frac{g_{r}\exp(-E_{r}/k_BT)}{\sum_{i}g_{i}\exp(-E_{i}/k_BT)}=\frac{g_{r}\exp(-E_{r}/k_BT)}{Q(T,\Upsilon)},
\end{eqnarray}
where we define the \textit{partition function} of the system as
\begin{eqnarray}
\label{eq:part}  Q(T,\Upsilon) = \sum_{i}g_{i}\exp(-E_{i}/k_BT),
\end{eqnarray}
and $\Upsilon$ denotes any other parameters that might govern the
values of $E_r$. Note that the summation goes over all energy levels
of the system and $g_i$ denotes the degeneracy of the state $E_i$,
i.e. the number of different states associated with the
energy level $E_i$. Thus, one may write $g_i=\Omega(E_{i})$, where
$\Omega$ comes from the previous formulation. Assuming the total
energy of the system to be an average energy of the
different states, i.e.
\begin{eqnarray}
\label{eq:engprobav}  E =\sum_{r}P_r E_r,
\end{eqnarray}
one can show that the Helmholtz free energy $\mathcal{F}$ can be written as
\begin{eqnarray}
\label{eq:Helm}  \mathcal{F} = -k_BT\ln Q.
\end{eqnarray}
Equation (\ref{eq:Helm}) provides the basic relation in the
canonical ensemble theory. Once $\mathcal{F}$ is known the other
thermodynamic quantities can be easily obtained.

Note that we have chosen the average energy to be the energy of the
system in this theory. One can show the total energy that we
associate to the system on micro canonical ensemble theory
corresponds to the most probable energy of the system, i.e. the
energy level that maximizes $P_r$ at a given temperature $T$. In
practice, i.e. in the thermodynamical limit
$N\longrightarrow\infty$, it can
be shown that these energies are equal and thus these two smilingly
different approaches are the same.

Finally, note that
\begin{eqnarray}
\label{eq:probHelm}  P_r =
\frac{g_{r}\exp(-E_{r}/k_BT)}{Q(T,\Upsilon)}=\frac{\exp[-(E_{r}-k_BT\ln g_{r})/k_BT]}{Q(T,\Upsilon)}=\frac{\exp(-\mathcal{F}_{r}/k_BT)}{Q(T,\Upsilon)},
\end{eqnarray}
where we use $S=k_B\ln \Omega$, which is justified by the
equivalence of the two ensemble theories. Equation
(\ref{eq:probHelm}) shows that to maximize $P_r$ at a fixed
temperature, we need to minimize $\mathcal{F}_r$ over
all admissible states $r$. To summarize, we have shown that
minimizing the Helmholtz free energy at a temperature $T$ (and
constant volume) is equivalent to finding the most probable energy
level, which is the total energy of the system. Note that this
minimization should be done over all variables that
determine the free energy.

\newpage
\bibliographystyle{elsarticle-num}
\bibliography{<your-bib-database>}





\end{document}